\documentclass[prb,aps,showpacs]{revtex4}

\usepackage[dvips]{graphicx}

\begin{document}


\bibliographystyle{prsty}

\author{Baruch Horovitz}

\affiliation{Department of Physics, Ben Gurion university, Beer Sheva
84105 Israel}

\author{Pierre Le Doussal}

\affiliation{CNRS-Laboratoire de Physique Th{\'e}orique de
l'Ecole Normale Sup{\'e}rieure, 24 rue Lhomond,75231 Cedex 05,
Paris France.}

\title{Disorder Induced Transitions in Layered Coulomb Gases
and Application to Flux Lattices in Superconductors}

\begin{abstract}
A layered system of charges with logarithmic interaction parallel
to the layers and random dipoles in each layer is studied via a
variational method and an energy rationale. These methods
reproduce the known phase diagram for a single layer where
charges unbind by increasing either temperature or disorder, as
well as a freezing first order transition within the ordered phase.
Increasing interlayer coupling leads to successive transitions in
which charge rods correlated in $N>1$ neighboring layers are
unbounded by weaker disorder. Increasing disorder leads to transitions
 between different N phases. The method is applied to flux
lattices in layered superconductors in the limit of vanishing
Josephson coupling. The unbinding charges are point defects in the
flux lattice, i.e. vacancies or interstitials. We show that short
range disorder generates random dipoles for these defects. We predict
and accurately locate a disorder-induced defect-unbinding
transition with loss of superconducting order, upon increase of
disorder. While $N=1$ charges dominate for most system parameters,
we propose that in multi-layer superconductors defect rods can be
realized.
\end{abstract}
\pacs{64.60.Cn,74.25.Qt,74.78.Fk}

\maketitle


\widetext




\section{Introduction}

There is considerable current interest in topological phase
transitions induced by quenched disorder, a problem relevant for
numerous physical systems. Such transitions are likely to shape
the phase diagram of type II superconductors. It was proposed
\cite{tgpldbragg} that the flux lattice (FL) remains topologically
ordered in a Bragg glass (BrG) phase at low field, and becomes
unstable to the proliferation of dislocations above some threshold
disorder (or field). The increased effect of disorder may lead to
increased critical current, this providing one scenario for the
ubiquitous and controversial "second peak" \cite{Kes,speakexp}
line in the phase diagram. Another scenario was proposed recently
\cite{H2} and is based on a disorder-induced decoupling transition
(DT) associated with the loss of superconducting order,
responsible for a sharp drop in the FL tilt modulus. An important
question then is whether this DT occurs before the BrG instability
(i.e within the BrG phase) or whether both occur simultaneously.

Theoretically, two types of phase transitions were shown to be
specific for pure layered superconductors. The first is decoupling
\cite{GK,Daemen,H1} at which the Josephson coupling as well as the
critical current between layers vanishes. The second is the
proliferation of point "pancake" vortices, vacancies and
interstitials (VI) in the FL above a temperature $T_{def}$ which,
above some field, is distinct from melting, as shown in the absence
of Josephson coupling \cite{Dodgson}. It is believed that this
pure system topological transition merges with the decoupling
transition \cite{Daemen,H1} as the bare Josephson coupling is
increased, being two anisotropic limits of the same transition
\cite{H3}. This transition induces a loss of superconducting order
(parallel to the layers by VI and perpendicular to them by
the layer decoupling) while the positional correlations of the
pure flux lattice is maintained \cite{FNF}. This transition has also
been studied in both limits in presence of
point impurity disorder \cite{H1,HL} as well as columnar disorder
\cite{MHL}. In
particular, we have recently demonstrated
\cite{HL} the existence of disorder-induced, VI unbinding
transition with loss of superconductivity in 3-dimensional (3D)
layered superconductors, which would be particularly relevant to
many layered superconductors and multilayer systems
\cite{Kes,Bruynseraede}.

Topological phase transitions in two dimensional systems
are conveniently studied using mapping onto Coulomb gases of charges
interacting via a long range logarithmic potential.
Studying general three dimensional systems, even for pure systems,
is considerably more difficult. The limit of layered superconductors with
magnetic coupling only, provides one rare example
where the problem can be studied analytically in 3D in
a controlled way. Indeed in this limit the
problem amounts to coupled layers with 2D Coulomb
interactions. In the presence of quenched disorder, the problem
becomes quite subtle already in 2D because charges
can freeze into inhomogeneous configurations. Progress
was made recently and it was shown
\cite{nattermann95,scheidl97,tang96,dcpld}
that quenched random dipoles lead to a phase transition, via
proliferation of defects at a finite threshold value of disorder,
even at temperature $T=0$. New analytical methods, based on
RG for the charge fugacity probability distribution,
and mapping onto a solvable model of directed polymer on the Cayley
tree were developped in 2D \cite{tang96,dcpld}.
In a short account of the present work \cite{HL}
we have extended some of these techniques to study the 3D system in presence of
disorder. Although a complete RG study along the lines
of \cite{dcpld} is possible in principle, we have
used simpler, and we believe largely equivalent, methods.
The first is an energy rationale which generalizes the
Cayley tree mapping. Second, we have introduced \cite{HL}
a novel Gaussian variational method which incorporates the
effect of the broad fugacity distribution, a feature
previously revealed by the RG \cite{scheidl97,dcpld}.
This method was also applied to the single and two layer case
in a related work on a random Dirac model relevant
to quantum Hall systems \cite{rdirac}.

The aim of this paper is to present details of our previous Letter
\cite{HL} focusing on two themes. First we consider a general
Disordered Coupled Coulomb Gas (DCCG) model system defined by
integer $\pm 1$ charges on $M$ layers in which the interaction

energy between two charges on layers $n$ and $n'$ is $2
J_{n-n'}\ln r$ with $r$ the charge separation parallel to the
layers; in addition the charges couple to quenched random dipoles.
A general study of this system is performed both via an energy
rationale and by a variational method, with consistent results.
These methods are explained in detail. Second, we apply this study
to various physical situations, mainly to layered superconductors
in an external field. We justify, stating clearly the assumptions,
that VI in the vortex lattice of layered superconductors with no
Josephson coupling and in the presence of pinning disorder can be
described by the DCCG model with quenched random dipoles.

In section II we present the DCCG model and its mapping to a
sine-Gordon type problem. In section III we develop a $T=0$ energy
rationale by an approximate mapping to Cayley tree problem. For
the one layer case we find the well known critical disorder value
of $\sigma_{cr}=1/8$ for the onset of VI. For the many layer case
we find that as the anisotropy $\eta =-J_1 /J_0$ increases a
cascade of phase transitions appear at which the number of
correlated charges on $N$ neighboring layers increases. These
"rod" phases appear at an decreasing critical disorder value until
at $\eta \rightarrow 1/2$ we find $N\rightarrow \infty$ and
$\sigma_{cr}\rightarrow 0$. In section IV we develop an efficient
variational method which is tested on the one layer system,
allowing for fugacity distributions, known \cite{dcpld} to be
important in 2D since disorder becomes broad at low temperature.
We reproduce the phase boundary in disorder-temperature plane
separating an ordered phase (bound charges) and a disordered
(unbound charges, i.e. finite VI density); the critical disorder
parameter at $T=0$ is $\sigma_{cr}=1/8$ is recovered. We also find
a first order line within the ordered phase (seen in the dynamics
study \cite{dcpld}) which becomes a crossover line in the
disordered phase. In section V we extend our method to the 2-layer
system and find for the anisotropy $\eta$ a critical value $\eta_c
=1-1/\sqrt{2}$  above which the single layer type transition is
preempted by a transition induced by bound states of two vortices
on the two layers with $\sigma_{cr}<1/8$, in agreement with the
energy rationale of section III. However, in a limited range of
$1-1/\sqrt{2}<\eta<1/3$ we find coexistence with a two gap state,
which is not captured by the energy rational in its simplest form,
but does not change the value of $\sigma_{cr}$. Of course, all of
these above results truly involve renormalized values of coupling
$J_n^ {ren}$ and disorder $\sigma^{ren}$. Although we have not
attempted a full RG study, one main additional effect of RG is
simply to substitute bare by renormalized values accounting for
screening effects, which on the basis of the two layer case can be
assumed to be small for our present purpose (i.e. identifying
transition lines at low temperature).

In section VI we develop the effective theory of layered
superconductors with magnetic coupling between layers, but without
Josephson coupling. We show that point disorder for the FL leads
to quenched dipoles for the VI, hence the DCCG problem. For typical
layered superconductors we predict the one layer type transition
with an effective disorder parameter. However, by increasing the
separation between layers, as in multilayer systems
\cite{Kes,Bruynseraede} to exceed the lattice parameter of the FL,
one may realize the new $N>1$ rod phases.

\section{Model for disordered layered Coulomb gas}

In this Section we define the model for $M$ coupled layers of
disordered Coulomb gases and also in terms of an equivalent
sine-Gordon model. Consider $n({\bf r},l)$ integer charges on the
$l$-th layer at position ${\bf r}$ within the layer. The
two-dimensional (2D) position ${\bf r}$ is defined on a lattice of
spacing $\xi$, which for the superconducting system is the
coherence length. We study the Hamiltonian:

\begin{eqnarray}
&& {\cal H}= - \frac{1}{2} \sum_{{\bf r} \neq {\bf r}'}\sum_{l,l'}
2 J_{ll'} n({\bf r},l) G({\bf r}-{\bf r}') n({\bf r}',l') -
\sum_{{\bf r},l} V_l({\bf r}) n({\bf r},l) + E_c\sum_{{\bf r},l}
n^2({\bf r},l) \label{H}
\end{eqnarray}
where $E_c$ is the core energy, accounting for short scale
energies $r < \xi$. Charges on the same or different layers interact with a
2D Coulomb interaction, with
\begin{eqnarray}
G({\bf r}) \approx_{|{\bf r}| \to \infty} \ln\frac{|{\bf
r}|}{\xi} \qquad G_{\bf q} \approx_{q \to 0} \frac{2 \pi}{q^2}
\end{eqnarray}
with $G({\bf r})=\int_{\bf q} G_{{\bf q}} (1- e^{i {\bf q}.{\bf
r}})$ and $\int_{\bf q} = \int \frac{d^2 q}{(2 \pi)^2}$ (on a square
lattice $G_{{\bf q}}^{-1}= \frac{1}{\pi} [2-\cos (q_{x}\xi)-\cos
(q_{y} \xi)]$). Neutrality is assumed in each layer. The disorder
potential $V_l({\bf r})$ can be considered as due to random
dipoles. A dipole has a potential $\sim 1/r$ or $\sim 1/q$ in
Fourier space; hence the disorder potential on the $l$-th layer
$V_l({\bf r})$ has long range correlations:
\begin{eqnarray}
&& \overline{V_l({\bf q}) V_{l'}({\bf q}')}
\approx_{q \to 0} 2 \Delta_{l l'} \frac{2 \pi}{q^2} (2 \pi)^2 \delta^{(2)}({\bf
q} +
{\bf q}' ) \\
&& \overline{(V_l({\bf r}) - V_l({\bf r}'))((V_{l'}({\bf r}) -
V_{l'}({\bf r}'))} \approx_{|{\bf r}- {\bf r}'| \to \infty} 4
\Delta_{ll'} \ln\frac{|{\bf r}-{\bf r}'|}{\xi}
\end{eqnarray}
where $\Delta_{ll'}\geq 0$. This logarithmically correlated
disorder is the one which exhibits a phase transition - other
types of disorder with either weaker or stronger correlations
result in either ordered or disordered phases, respectively, hence
no phase transition as function of the disorder strength. One
simpler case, which we will study in details, is the case of
uncorrelated disorder from plane to plane, namely $\Delta_{l l'} =
\sigma J_0^2 \delta_{ll'}$. In that case one has
\begin{eqnarray}
 \overline{[V_l({\bf r})-V_{l}({\bf r}')]^2} \approx_{|{\bf r} - {\bf r}'|
\to \infty} 4 \sigma J_0^2
 \ln\frac{|{\bf r}-{\bf r}'|}{\xi} \label{corr}
\end{eqnarray}
It is clear that the model on a square lattice defined by its
partition sum $Z_{\text{latt}}  = \sum_{\{ n({\bf r},l) \}} e^{-
\beta H}$ can also be seen as a neutral 2D Coulomb gas model for
$M$-component vector charges. A given configuration of charges is
thus defined by a set of vector charges $\{n({\bf
r},l)\}_{l=1,..M}$ on a 2D lattice.

We define the Fourier transform:
\begin{eqnarray}
n({\bf q},k) =  d \sum_l \sum_{\bf r} n({\bf r},l) e^{i {\bf q}\cdot
{\bf r} + i k d l}
\end{eqnarray}
where d is the spacing between layers and in a continuum limit $d
\sum_l\rightarrow\int dz$ and  $\xi^2 \sum_{\bf}\rightarrow\int
d^2r$. The inverse formula for the charge density (per unit area)
is
\begin{eqnarray}
n({\bf r},l)/\xi^2 = \int_k \int_{\bf q} n({\bf q},k) e^{-i {\bf
q}\cdot {\bf r}- i k d l}
\end{eqnarray}
with $\int_{\bf q} = \int \frac{d^2 q}{(2 \pi)^2}$, $k=2 \pi
{\tilde m} /M d$ with ${\tilde m}$ integer, $-[M/2]+1 \leq {\tilde
m} \leq [M/2]$, and $\int_k = \frac{1}{M d} \sum_k \to \int_{-
\pi/d}^{\pi/d} \frac{d k}{2 \pi}$ at large $M$.

We perform disorder averages via the replica method, i.e. from the
replicated partition function $\overline{Z^{m}}$ in the limit $m
\to 0$,  disorder averaged correlations and free energy are
obtained. For integer $m$ we have
\begin{equation}\label{Z0}
\overline{Z^m}=\sum_{\{n_a({\bf r},l)\}} e^{- \beta H^{(m)}}
\end{equation}
with $\beta = 1/T$, which on a lattice is {\it exactly} a Mm-component 2D
vector Coulomb gas
  with integer charges at
each site ${\bf r}$ with integer entries $n_{a}({\bf r},l)$ at each $a=1,..m$,
$l=1,..M$. The
replicated Hamiltonian is \cite{footnote0}
\begin{eqnarray}  \label{latticereplica}
&&  \beta H^{(m)} = - \sum_{{\bf r} \neq {\bf r}'} K_{la,l'b}
~n_{a}({\bf r},l) G({\bf r}-{\bf r}') n_{b}({\bf
r}',l') + \beta E_c \sum_{{\bf r},l,a} n^2_a({\bf r},l)\\
&& K_{la, l'b}=\beta J_{ll'} \delta_{ab} - \beta^2 \Delta_{ll'}
\end{eqnarray}
summation over repeated indices is assumed unless otherwise specified.

For system which is cyclic and (statistically) translationally
invariant in the $z$ direction, i.e:
\begin{eqnarray}
J_{ll'}=J_{|l- l'|} \qquad \Delta_{ll'}=\Delta_{|l- l'|}
\end{eqnarray}
it is convenient to work with a Fourier space version which
reads:

\begin{eqnarray}
&& \beta {\cal H}^{(m)}= \frac{1}{2 d^2}  \int_k \int_q n_a({\bf
q},k) (G_0)_{ab}({\bf q},k) n_b^*({\bf q},k)
+ \beta E_c \sum_{{\bf r},l,a} n^2_a({\bf r},l) \label{Hm} \\
&& (G_0)_{ab}({\bf q},k)= \frac{4\pi}{q^2} [g(k) \delta_{a,b} -
\beta^2 \Delta(k)] \label{G0}
\end{eqnarray}
 For later
convenience we have defined  $g(k)=\beta J(k)$, $J(k)= d \sum_l
J_l \exp(i k d l)$, with $J_l = \int_k J(k) \exp {- i k d l}$.
Similarly  $\Delta(k)= d \sum_l \Delta_{l,0} \exp(i k d l)$, i.e.
for disorder uncorrelated between layers $\Delta(k) = d \sigma
J_0^2$.

We proceed to define an equivalent sine-Gordon system. We first
rewrite the logarithmic interaction by using a scalar field $\chi
_a({\bf r},l)$,
\begin{equation}\label{Z1}
\overline{Z^m}=\langle \sum_{\{n_a({\bf r},l)\}}\prod_{{\bf
r},l,a}e^{-i\chi _a({\bf r},l)n _a({\bf r},l)- \beta E_c\sum_{{\bf
r},l,a} n^2_a({\bf r},l)}\rangle_{\chi}
\end{equation}
The average is done with the weight $\exp[- \frac{1}{2} \int_k
\int_q \chi _a({\bf q},k) (G_0^{-1})_{ab}({\bf q},k) \chi^*_b({\bf
q},k)]$; performing this Gaussian average one readily recovers Eq.
({\ref{Z0}). The inverse of Eq. (\ref{G0}) is derived by the
inversion formula $(A \delta_{ab}+B)^{-1}=(1/A)\delta_{ab}- B
/[A(A+B m)]$, which for $m \to 0$ yields
\begin{equation}
(G_0^{-1})_{ab}({\bf q},k) = \frac{q^2}{4\pi} [\frac{1}{g(k)} \delta_{a,b} +
\beta^2 \frac{\Delta(k)}{g^2(k)} ] \,.\label{G1}
\end{equation}
The product in Eq. (\ref{Z1}) at each lattice point ${\bf r}$ can be
written as a sum of all $\pm 1,0$ values of $n_a({\bf r},l)$, i.e.
a sum on all integer vector ${\bf n}=\{n_{a,l}\}$; $a=1,..m$,
$l=1,..M$
\begin{equation}
\overline{Z^m}=\langle \prod_{{\bf r}}[1+\sum_{\{n\neq
0\}}Y[{\bf n}]e^{i\sum_{a,l}n_{a,l}\chi_a({\bf
r},l)}]\rangle_{\chi}\label{Z2}
\end{equation}
where the fugacity is $Y[{\bf n}]=\exp[- \beta E_c\sum_{a,l} n^2_{a,l}]$. At
this point we make an approximation of small fugacities $Y[{\bf n}]$ (dilute
limit)
and write the above as an exponent
\begin{equation}\label{Z3}
\overline{Z^m}=\langle \prod_{{\bf r}}\exp[\sum_{\{n\neq
0\}}Y[{\bf n}]e^{i\sum_{a,l} n_{a,l}\chi_a({\bf r},l)}]\rangle_{\chi}\
\,.
\end{equation}
This approximation neglects harmonics of $\exp[{\bf n}\cdot {\bf \chi}]$,
 i.e. it neglects vector charges with
entries $|n_{a,l}|>1$. These harmonics are irrelevant near the actual
 phase transition \cite{dcpld}.
 Here and below we define ${\bf n} \cdot {\bf \chi}  =
\sum_{l a} n_{a,l} \chi_{a}({\bf r},l)$.
The result Eq. (\ref{Z3}) can now be
identified as the partition sum for a sine-Gordon type
Hamiltonian,
\begin{equation}\label{SG}
\beta {\cal H}_{SG} = \frac{1}{2} \int_k \int_q \chi _a({\bf q},k)
(G_0^{-1})_{ab}({\bf q},k) \chi^*_b({\bf q},k) -\sum_{{\bf r}}
\sum_{{\bf n}} Y[{\bf n}] \exp{i {\bf n} \cdot {\mbox{\boldmath
$\chi$}}({\bf r})}\,.
\end{equation}
where $\chi _a({\bf q},k) = \xi^2 d \sum_{{\bf r}} \sum_l \chi
_a({\bf r},l) e^{i k d l + i {\bf q} \cdot {\bf r}}$. We note that
the $+$ sign for the off diagonal replica term in Eq. (\ref{G1})
 corresponds to imaginary gauge disorder in a related Dirac problem
 \cite{rdirac}. The
validity of the approximations leading to (\ref{SG}) are discussed
in \cite{dcpld} in the context of a single layer. As also shown
below, it is important, as done here, to retain replica charges
with several non zero entries in order to describe the freezing
transitions at low temperatures.

\section{energy rationale}

In this section we consider the Coulomb gas problem at $T=0$ and
develop an energy rationale to determine the phase diagram of the
coupled layer system. The problem amounts to find minimal energy
configurations of charges in a logarithmically correlated random
potential. To ascertain the XY ordered phases (bound defects)
and the transitions out of it (defect unbinding), a first step is to study the
dilute limit of a single charge (or dipole). Even then, the
full analytical solution is difficult, but various approximations
have been argued to give exact leading order results.
For a single layer it was studied either using
\cite{nattermann95,korshunov_nattermann_diagphas,castillo,rem} a ``random
energy model'' (REM)
approximation, or more accurately using a
representation in terms of directed polymers on a Cayley tree
(DPCT), introduced in \cite{tang96,chamon96}. The continuous version of the
DPCT representation (branching process) was shown to emerge
\cite{dcpld} from the one loop Coulomb gas RG of the single layer
problem, both for the single charge (or dipole) problem and
for the many charges problem including screening effects. It is
thus expected to be accurate.

Schematically, one considers a tree with independent random
potentials (Fig. 1 inset) $v_i$ on each bond with variance ${\bar
v_i^2}=2 \sigma J_0^2$. For definiteness we can discuss a tree of
coordination $e^2$, the choice being immaterial for our present
considerations. After $p$ generations one has $\sim e^{2 p}$ sites
which are mapped onto a 2D layer: each point ${\bf r}$ corresponds
to a unique path on the tree with $v_1,...,v_p$ potentials and is
assigned a potential $V({\bf r})=v_1+...+v_p$. Two points ${\bf
r}$, ${\bf r}'$, separated by $|{\bf r} - {\bf r}'| \sim e^{p'} $
in Euclidean space, have a have a common ancestor at the previous
$p' \approx \ln |{\bf r} - {\bf r}'|$ generation Since all bonds
previous to the common ancestor are identical $\overline{[V({\bf
r})-V({\bf r}')]^2}=2\sum_{i=1}^{p}{\bar v_i^2} = 4\sigma J_0^2
\ln (|{\bf r}-{\bf r}'|)$, reproducing Eq.\ (\ref{corr}) on each
layer. Thus the growth of correlations on the tree and in
Euclidean space is by construction the same, and the single charge
problem corresponds to a single directed polymer. Exact solution
of  DPCT \cite{ct} yields the best energy gained from disorder
$V_{min}=min_{\bf r} V({\bf r}) \approx - \sqrt{8 \sigma} J_0 \ln
L$ for a volume $L^2$, with only $O(1)$ fluctuations \cite{dcpld},
i.e $- \sqrt{8 \sigma} J_0$ per generation $p=\ln L$. It is argued
that this is also the exact result for the Euclidean problem. For
a dipole in a single layer, one consider two directed polymers on
the same Cayley tree. Opposite sign charge see opposite disorder
$-v_i$, the gain from disorder $- V_{max}$ behaving identically.
The configurations of the two oppositely charged polymers can
however being argued to be essentially independent (i.e.
determining maximum and minimum of a log-correlated landscape can
be performed independently).

To generalize the Cayley tree argument we construct optimal energy charge
configurations for $M$ coupled layers as follows. Consider $N$
neighboring layers with a $+,-$ pair on each layer and no charges
on the other $M-N$ layers. We assume, for convenience, that
$J_0>0$ and $J_{l \neq 0} \leq 0$ so that {\em equal} charges on
different layers attract. The DPCT representation now involves on
a single tree $N$ $+$ polymers (each seeing different disorder)
and $N$ $-$ polymers (each seeing opposite disorder $-v_i$ to
their $+$ partner). A plausible configuration is that the $+$
charges bind within a scale $L^{\epsilon}$ ($0 \le \epsilon \le
1$), so do the $-$ charges, while the $+$ to $-$ charge
separations define the scale $L$. Its tree representation (Fig.
1a) has $2N$ branches with $\epsilon \ln L$ generations, i.e. an
optimal energy of $-2 N \sqrt{8\sigma}J_0\epsilon \ln L$. On the
scale between $L^{\epsilon}$ and $L$ the $+$ charges act as a
single charge with a potential $\sum_{l=1}^N V_l({\bf r})$ (the
$N$ polymers share the same branch) of variance $N\sigma$ hence
the optimal energy is $-2\sqrt{8N\sigma}J_0 (1-\epsilon)\ln L$.
Note that the rod formation limits the disorder optimization
leading to a disorder energy $\sim \sqrt{N}<N$. The total energy
gain from the disorder potential is thus estimated as:
\begin{equation}
E_{dis} \approx -2J_0\sqrt{8\sigma}[\epsilon N
+(1-\epsilon)\sqrt{N}]\ln L \,.
\end{equation}

It is clearly exact for both $\epsilon=0$ and $\epsilon=1$,
sufficient for our purpose. This result can also be obtained from
the REM approximation, i.e. replacing the $V({\bf r})$ by $L^2$
variables {\it uncorrelated} in ${\bf r}$, with the same on-site
variance $\overline{V^2({\bf r})} \sim 2 \sigma J_0^2 \ln L$ also
yielding \cite{rem} $V_{min} \sim - \sqrt{8 \sigma} J_0 \ln L$.

The competing interaction energy from the couplings $J_l$ is for
the $+-$ pairs $[2J_0N+4 \sum_{l=1}^{N}J_l(N-l)]\ln L$ while for the
$++$ and $--$ pairs it is $-4\sum_{l=1}^N J_l(N-l)\epsilon \ln L$.
Hence the interaction energy is
\begin{equation}
E_{int} = 2J_0N[1-2\sum_{l=1}^N \eta _l (1-l/N)(1-\epsilon)]\ln L
\end{equation}
where $\eta _l =-J_l/J_0$. The total energy
$E_{tot}=E_{dis}+E_{int}$ is linear in $\epsilon$, hence the
minimum is at either $\epsilon=1$ or at $\epsilon=0$. Since
$\epsilon =1$ implies that the $+$ charges unbind, it is
sufficient to consider $\epsilon=0$ with all $N \geq 1$, i.e. a
rod is aligned with $N$ correlated charges at distance $O(1)$ and
has energy
\begin{equation}
E_{tot} = 2J_0N[1-2\sum_{l=1}^N \eta_l(1-\frac{l}{N})
-\sqrt{\frac{8\sigma}{N}}]
\ln L \,. \label{Etot}
\end{equation}
One can introduce more scales $L^{\epsilon '}$ to describe the
multi charges, however, as the energy is linear in $\epsilon '$
the result is the same rod structure.

\begin{figure}[htb]
\begin{center}
\includegraphics[scale=0.5]{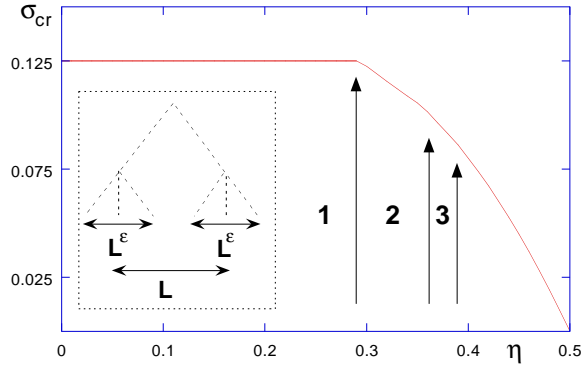}
\end{center}
\caption{Critical disorder values with only nearest neighbor
coupling $J_1$ vs. the anisotropy $\eta=-J_1/J_0$. Transitions
between different $N$ phases are marked with arrows. Inset: the
Cayley tree representation (for $N=3$ neighboring layers) with $+$
charges (at the tree endpoints) separated by $L^\epsilon$ along
the layers, and separated by $L$ from the $N=3$ $-$ charges.}
\label{fig1}
\end{figure}

Consider first the case with only nearest neighbor coupling $\eta
_1$ and only intralayer disorder correlation $\sigma J_0^2$.
Disorder induces the $N$ vortex state (i.e. $E_{tot}$ vanishes) at
the critical value
\begin{equation}
\sigma_{cr}^{(N)}=\frac{N}{8}[1-2\eta_1 (1-\frac{1}{N})]^2 \,.
\end{equation}
The system is thus fully stable to disorder only for $\sigma < \sigma_{cr}$
with:
\begin{equation}
\sigma_{cr} = \min_N \frac{N}{8}[1-2\eta_1 (1-\frac{1}{N})]^2 \,.
\end{equation}
When $\sigma$ reaches $\sigma_{cr}$ the first instability is to
one of the $N$ rod state, where $N$ depends on the value of the
anisotropy $\eta_1$. If $\eta _1 < \eta _1^{(1)} = 1-1/\sqrt{2}$
then $\sigma_{cr}^{(N)}$ is minimal at $N=1$ and the first
instability is similar to the one of a single layer with $\sigma_
{cr}=1/8$. For larger anisotropies $\eta _1^{(N-1)} <  \eta _1 <
\eta _1^{(N)}$, the first instability occurs at $\sigma_{cr} =
\sigma_{cr}^{(N)}$ towards a $N$-rod state with $1/(1-2\eta
_1^{(N)}) = 1 + \sqrt{N(N+1)} \sim N$, thus with diverging $N$ as
$\eta _1^{(N)} \to \frac{1}{2}$ (Fig 1) (for $\eta _1>\frac{1}{2}$
$E_{tot}<0$ even without disorder and the defects would form a
lattice at $T=0$).

Upon increasing $\sigma$ beyond $\sigma_{cr}^{(N)}$ a given rod
phase $N>1$ would eventually decompose into the $N=1$ phase. In
particular the energies of the $N=1,2$ phases become equal at
$\sigma_{cr}^{(1,2)}=\eta_1^2/[4(\sqrt{2}-1)^2]$ which equals 1/8
at $\eta_1=1-1/\sqrt{2}$. Hence at $\eta_1>1-1/\sqrt{2}$ the N=2
rods disintegrate into N=1 charges at
$\sigma>\sigma_{cr}^{(1,2)}$. The variational solution (section
VB) shows that this secondary line is actually at a somewhat lower
$\sigma_{cr}^{(1,2)}$ (see Fig. 5 below).

In the general case with all couplings $J_l$ the critical value
is:
\begin{equation}
\sigma_{cr}^{(N)}=\frac{N}{8}[1-2\sum_{l=1}^N\eta_l (1-\frac{l}{N})]^2
\,.
\end{equation}

We consider in particular $J_l$ with range of $l_0$ constrained by
$\sum_l J_l=0$, as relevant for the superconductor system (section
VI). An illustrative example is $\eta_l=\eta_1 \exp (-(l-1)/l_0)$,
constrained as $\eta_1 = \frac{1}{2} (1 - \exp(-1/l_0))$ (note
that $\sum_{l=1}^N \eta_l = \frac{1}{2} (1 - \exp(-N/l_0))$. One
then has:
\begin{equation}
\sigma_{cr}^{(N)}=\frac{1}{8 N}
\frac{(1-\exp(-N/l_0))^2}{(1-\exp(-1/l_0))^2}
\end{equation}
For large $l_0 \gg 1$, each $\eta _{l\ne 0}$ is small:
$\sigma_{cr}^{(N)}$ as a function of $N$ starts by increasing and for $N
\lesssim l_0$ the lowest $\sigma_{cr}^{(N)}$ is at $N=1$. However, the
combined strength of $N \approx l_0$ vortices being significant,
it has a maximum and then decreases back to zero for $N > l_0$ as
$\sigma_{cr}^{(N)} \approx l_0^2/8N$. Hence $\sigma_{cr}^{(N)} \rightarrow 0$
as $N \rightarrow \infty$ and any small disorder seems to nucleate
such vortices. This is because of the perfect screening of the
zero mode $\sum_l J_l=0$ which implies that an infinite charge rod
has a vanishing $\ln r$ interaction; hence a logarithmically
correlated disorder is always dominant.

In practice, the realization of these large $N$ states depends,
however, on the type of thermodynamic limit. Adding to Eq.\
(\ref{Etot}) the core energy $2 E_c N$ yields
\begin{equation}
E_{tot}' = 2 J_0 \sqrt{N} (\sqrt{8\sigma_{cr}^{(N)}} -
\sqrt{8\sigma}) \ln L + 2 E_c N
\end{equation}
which becomes negative only beyond the scale
\begin{equation}
L_N \approx \exp
\{E_c\sqrt{N}/[J_0(\sqrt{8\sigma}-\sqrt{8\sigma_{cr}^{(N)}})]\}
\label{L}\,.
\end{equation}
This $L_N$ is the typical distance between rod vortices. Hence
even if $\sigma>\sigma_{cr}^{(N)}$ only for system size $L>L_N$
the energy gain from disorder wins over core (and interaction)
energy. Hence as $\sigma\rightarrow 0$ such states are only
achievable in a thermodynamic limit where $L/N$ diverges
exponentially. Using $\sigma_{cr}^{(N)} \approx l_0^2/8N$, for $N>
l_0^2/8\sigma$ the lowest scale $L$ in this range is achieved at
$N=l_0^2/2\sigma$ and leads to a (system size) lower bound
$L_{min}\approx \exp [E_c l_0/4J_0 \sigma]$ for observing large
$N$ states with a given $\sigma <\frac{1}{8}$. For layered
superconductors \cite{footnote1} $E_c/J_0 \gg 1$ and $l_0 \gg 1$
and this large $N$ instability occurs at unattainable scales, thus
$N=1$ dominates. One needs $l_0\approx 2-3$ to realize the $N>1$
states, attainable in multilayers (see discussion section VII).

We finally generalize the energy argument for the $\epsilon =0$
configuration to the case of arbitrary correlations $\gamma_l =
\Delta_l/\Delta_0$. The disorder energy can be found from the
variance of $\sum_{i=1,N} V_l({\bf r})$ leading to the replacement
$\sigma \to \sigma (1 + 2 \sum_{l=1}^N (\Delta_{l,0}/\Delta_0)
(1-l/N))$ in Eq. (\ref{Etot}). A more compact form can be obtained
by writing directly
\begin{equation}
E_{tot} = 2 [ \sum_{l=1}^N \sum_{l'=1}^N J_{ll'} - \sqrt{8
\sum_{l=1}^N \sum_{l=1}^N \Delta_{ll'}} ]\ln L
\end{equation}
Using that:
\begin{eqnarray}
&& \sum_{l=1}^N \sum_{l'=1}^N J_{ll'} = \int_k J(k) \phi_N(k) \\
&& \phi_N(k) =\sum_{ll'}^N e^{ikd(l-l')} =\frac{\sin^2(N
kd/2)}{\sin^2(kd/2)} \label{phiN}
\end{eqnarray}
One has the criticality condition for a $N$ rod:
\begin{eqnarray}\label{Nrod}
\int_k \Delta(k) \phi_N(k) = \frac{1}{8} (\int_k J(k) \phi_N(k))^2
\end{eqnarray}
which in terms of $\sigma=\Delta_0/J_0^2$ has the critical value
\begin{eqnarray}
\sigma_{cr}^{(N)} = \frac{1}{8} \frac{ (\int_k \frac{\sin^2(N
kd/2)}{\sin^2(kd/2)} J(k)/J_0)^2}{ \int_k \frac{\sin^2(N
kd/2)}{\sin^2(kd/2)} \Delta(k)/\Delta_0}
\end{eqnarray}
For fixed anisotropies $J(k)/J_0$, $\Delta(k)/\Delta_0$ this
relates the overall critical disorder strength $\sigma_{cr}^{(N)}$ to the rod
length
$N$.

\section{variational method - the single layer}

We develop here a variational method which allows for fugacity
distributions, an essential feature in the one-layer problem. The
method is developed in this section for the one-layer system and
it is shown that one recovers in a simple way all the important
known features for this problem. Furthermore, new insight is
gained for a critical line within the ordered (charge bound)
phase, as well as a crossover line in the disordered (charge
unbound) phase, at which the the functional dependence of the
charge density changes.

The single layer replicated Coulomb gas Hamiltonian is
\begin{equation}\label{2d}
\beta {\cal H}^{(m)} = \frac{1}{2}  \int_q n_a({\bf
q})\frac{4\pi}{q^2} [K\delta_{ab}-\sigma K^2] n_b^*({\bf q}) +
\beta E_c\sum_{{\bf r},a} n^2_a({\bf r})  \,.
\end{equation}
where $n_a({\bf q})= \sum_{\bf r} n_a({\bf r}) e^{ i {\bf q} \cdot {\bf r} }$.
Note that $\sigma >0$ is here essential; the same 2D system with
$\sigma <0$ has been shown to have a different phase diagram
\cite{H4,scheidl98}. The equivalent sine-Gordon system is now
\begin{eqnarray}
&& \beta {\cal H}_{SG} = \frac{1}{2} \int_q \chi _a({\bf q})
(G_0^{-1})_{ab}(q) \chi^*_b({\bf q}) -\sum_{{\bf r}} \sum_{{\bf
n}} Y[{\bf n}] \exp{i {\bf n} \cdot
{\mbox{\boldmath $\chi$}}({\bf r})} \label{sg}  \\
&& (G_0^{-1})_{ab}(q) = \frac{q^2}{4 \pi} [\frac{1}{K} \delta_{a b} +
\sigma ]
\end{eqnarray}
where $\chi_a({\bf q}) = \xi^2 \sum_{\bf r} \chi_a({\bf r}) e^{ i {\bf q} \cdot
{\bf r} }$
and bare fugacities $Y[{\bf n}] = \exp(- \beta E_c\sum_a n_{a}^2 )$.
Here one has simply ${\bf n} \cdot {\mbox{\boldmath $\chi$}}({\bf
r}) = \sum_{a} n_{a} \chi_a ({\bf r})$, with
${\bf n}$ a nonzero vector with entries $n_a=\pm 1, 0$.

The variational method represents the full Hamiltonian (\ref{sg})
by an optimal Gaussian one of the form
\begin{eqnarray}
\beta {\cal H}_{var} = \frac{1}{2} \int_q G^{-1}_{ab}(q) \chi_a({\bf q}) \chi_b
(-{\bf q})
\end{eqnarray}
where $G_{ab}$ is to be determined by a variational principle. The
variational free energy is $F_{var} = F_0 + <{\cal H}_{SG} - {\cal H}_{var}>_
{{\cal H}_{var}}$
with $\beta F_0 = - \ln Z_0 = - \frac{1}{2} Tr \ln G$  is found to
read:
\begin{eqnarray}\label{Fvar}
\frac{\beta F_{var}}{L^2}  = - \frac{1}{2} \int_q Tr \ln G(q) +
\frac{1}{2} \int_q Tr( G_0^{-1}(q) G(q)) - \xi^{-2} \sum_{{\bf n}
\neq 0} Y[{\bf n}] e^{ - \frac{1}{2} \int_q {\bf n}\cdot G(q)\cdot
{\bf n}}
\end{eqnarray}
up to an unimportant constant, where the $Tr$ is in replica indices.
Taking the derivative $\frac{\delta}{\delta G_{ab}(q)}$ one
obtains the saddle point equation:
\begin{eqnarray}\label{Svar}
\sigma_{ab} = \xi^{-2} \sum_{{\bf n} \neq 0} n_a n_b Y[{\bf n}] e^{ -
\frac{1}{2} {\bf n}\cdot G\cdot {\bf n}}
\end{eqnarray}
where we have defined:
\begin{eqnarray}\label{Ginverse}
&& G^{-1}_{ab}(q) = (G_0)^{-1}_{ab}(q) + \sigma_{ab}
\end{eqnarray}

We recall first some technical relations \cite{scheidl97,dcpld}.
In the following we represent relevant operators as averages which
depend only on $n_{+},n_{-}$, which are the number of $+$ or $-$
entries in ${\bf n}$, respectively. The averages have the form
\begin{eqnarray}
A[{\bf n}] = \int dudv \Phi(u,v) e^{ u (n_{+} + n_{-}) + v (n_{+}
- n_{-})}
\end{eqnarray}
where $z_{\pm} = e^{u \pm v}$ can be interpreted as fugacities for the $\pm$
charges, hence $\Phi(u,v)$ is a fugacity distribution. A sum on
all ${\bf n}\neq 0$ can be written in terms of the variables
$n_+,n_-$ with a combinatorial factor for the number of ${\bf n}$
vectors with a given $n_+,n_-$,
\begin{eqnarray}\label{bin1}
&& \lim_{m \to 0} \frac{1}{m} \sum_{{\bf n} \neq 0} A[{\bf n}] =
\lim_{m \to 0} \frac{1}{m}\sum_{0< n_{+} + n_{-} \leq m}
\frac{m!}{n_+ ! n_- ! (m - n_+ - n_-)!}
\langle e^{ (u+v)n_{+}+(u-v)n_{-}}\rangle_\Phi \\
&& =\lim_{m \to 0}<(1 + e^{u+v}  + e^{u-v})^m/m >_{\Phi}=
 \langle\ln(1 + e^{u+v}  + e^{u-v}) \rangle_{\Phi}
\end{eqnarray}
and the binomial expansion has been used and $<...>_{\Phi}$
denotes an average with the weight $\Phi(u,v)$. Similarly one has:
\begin{eqnarray}\label{bin2}
&& \lim_{m \to 0}\frac{1}{m} \sum_{{\bf n} \neq 0} \sum_{a} n_a^2
A[{\bf n}] =\langle \sum_{{\bf n} \neq 0} (n_+ + n_-)
 e^{ u (n_{+} + n_{-}) + v (n_{+} - n_{-})} \rangle_{\Phi} \nonumber\\
&& = \langle \partial_u \ln(1 + e^{u+v}  + e^{u-v}) \rangle_{\Phi}
 = \langle \frac{e^{u+v} + e^{u-v}}{1 + e^{u+v}  + e^{u-v}} \rangle_{\Phi}
\end{eqnarray}
and
\begin{eqnarray}\label{bin3}
&& \lim_{m \to 0} \frac{1}{m}\sum_{{\bf n} \neq 0} \sum_{a,b} n_a
n_b A[{\bf n}] =\langle \sum_{{\bf n} \neq 0} (n_+ - n_-)^2
 e^{ u (n_{+} + n_{-}) + v (n_{+} - n_{-})} \rangle_{\Phi} \nonumber\\
&& = \langle \partial_v^2 \ln(1 + e^{u+v}  + e^{u-v})
\rangle_{\Phi} = \langle \frac{e^{u+v} + e^{u-v} + 4 e^{2 u}}{(1 +
e^{u+v}  + e^{u-v})^2} \rangle_{\Phi}
\end{eqnarray}

In our case we consider a replica symmetric parametrization
$\sigma_{ab} = \sigma_c \delta_{ab} + \sigma_0$ so that
$G_{ab}=\int_q G_{ab}(q)$ has the form $G_{ab} = G_c \delta_{ab} -
A$, where
\begin{eqnarray}
&& G_c = \int_q \frac{1}{\frac{q^2}{4 \pi K} + \sigma_c}=
K\ln\frac{\Lambda^2}{4\pi K\sigma_c}  \\
 && A= \int_q \frac{\sigma_0 + \frac{\sigma q^2}{4 \pi}}{(\frac{q^2}{4
\pi K} + \sigma_c)^2}=K^2\sigma \ln\frac{\Lambda^2}{4\pi
K\sigma_c}-K^2\sigma+\frac{K\sigma_0}{\sigma_c}\label{A}
\end{eqnarray}
where $\Lambda \sim \xi^{-1} \gg K\sigma_c$ is a cutoff on the q integration.
Since $\int_q {\bf n}\cdot G(q)\cdot{\bf n}= G_c \sum_a n_a^2 - G
(\sum_a n_a)^2$ we can now identify the weight function from the
interaction term in Eq. (\ref{Fvar}),
\begin{eqnarray}\label{int}
&& Y[{\bf n}] e^{ - \frac{1}{2} \int_q {\bf n}\cdot G(q)\cdot {\bf
n}}=e^{- (\frac{1}{2} G_c + \beta E_c) (n_+ + n_-) + \frac{1}{2} A (n_+ - n_-)
^2}\nonumber\\
&&=\int dv \Phi(v) e^{ u (n_{+} + n_{-}) + v (n_{+} - n_{-})}
\end{eqnarray}
where here the weight function depends here only on $v$
\begin{eqnarray}
\Phi(v) =  \frac{1}{\sqrt{2 \pi A}} e^{- v^2/(2 A)}
\label{proba}
\end{eqnarray}
and $u= - \beta E_c - \frac{1}{2} G_c$. We recall that:
\begin{eqnarray}
y = e^{ - \beta E_c}
\end{eqnarray}
is the bare fugacity of the charge, while the $z_{\pm}$
corresponds to the renormalized ones (they become
random variables because of the quenched disorder
in the system). The bare model can be generalized by
introducing short-ranged randomness in the bare
core energies (of width $E_0$) \cite{scheidl97,dcpld},
resulting in the replica symmetric form $Y[{\bf
n]}=\exp[- \beta E_c\sum_an_a^2 - \frac{1}{2} \beta^2  E_0^2 \sum_{ab} n_a n_b]
$.
This corresponds to the change $A \to A + \beta^2 E_0^2$ in the averages above.
Since A is divergent at criticality a finite $E_0$ can be ignored.

The interaction term in Eq. (\ref{Fvar}) is therefore
\begin{equation}\label{int2}
 \sum_{{\bf n}\neq 0} Y[{\bf n}] e^{ - \frac{1}{2} \int_q {\bf n}\cdot G(q)
\cdot {\bf
n}}=\langle \ln(1 + e^{u+v}  + e^{u-v}) \rangle_{\Phi}
\end{equation}

To identify $\sigma_c,\sigma_0$ we consider the variational
equation (\ref{Svar}) and note that Eq. (\ref{bin3}) in the limit
$m \to 0$ is $\sigma_c$, while Eq. (\ref{bin2}) is
$\sigma_c+\sigma_0$, hence
\begin{eqnarray}
&& \sigma_c = \xi^{-2}
\langle\frac{e^{u+v} + e^{u-v} + 4 e^{2 u}}{(1 + e^{u+v} + e^{u-v})^2}
\rangle_{\Phi}\label{sc}\\
&& \sigma_0 = \xi^{-2} \langle \frac{(e^{u+v} - e^{u-v})^2}{(1 + e^{u+v} +
e^{u-v})^2} \rangle_{\Phi}\label{s0}\,.
\end{eqnarray}
These equations, together with (\ref{proba}) and (\ref{A}) form
the closed set of self-consistent equations that we want to solve.
On general grounds one expects an ordered phase where the self
energy $\sigma_c$ vanish corresponding to zero charge density and
zero renormalized fugacity (XY phase). The solution with $\sigma_c
>0$ corresponds to a phase with finite density of charges
(disordered phase), the typical correlation length (see Eq.
\ref{Ginverse}) being $\sim \sigma_c^{-1/2}$,the typicall distance
between charges. We will thus perform the analysis near the
critical line, where $\sigma_c$ is small. We will first neglect
the $\sigma_0$ term in (\ref{A}) and later show that it is indeed
negligible in all regimes of interest.

To analyze these equations we note that the $v$ integration is
dominated by large $|u|$ and $A$ which diverge at criticality,
$\sigma_c \to 0$. The function displayed in (\ref{sc}) is maximal
at $v=-u$ with a width $O(1)$, while the gaussian $\Phi(v)$ is
maximal at $v=0$ with a width $O(\sqrt{A})$. Consider then
 $v>0$ where the $e^{u+v}$ term dominates and is either very small
($u+v<0$) or very large ($u+v>0$), hence
\begin{equation}\label{sc1}
\xi^2\sigma_c \approx
2\int_0^{\infty}\frac{e^{u+v-v^2/2A}}{1+e^{u+v}}\frac{dv}{\sqrt{2\pi
A}} \approx 2\int_0^{-u}e^{u+v-v^2/2A}\frac{dv}{\sqrt{2\pi A}}
+2\int_{-u}^{\infty}e^{-u-v-v^2/2A}\frac{dv}{\sqrt{2\pi A}}\,.
\end{equation}
In the second term the saddle point at $v=-A$ is outside of the
integration range, hence it is dominated by the lower limit, i.e.
it is of order $\exp(-u^2/2A)$. The first term has a saddle point
at $v=A$ which is within the integration range if $A<-u$ and then
\begin{equation}\label{sc2}
\sigma_c \sim e^{u+A/2}\sim y \sigma_c^{(K-K^2\sigma)/2} \qquad
A<-u \,.
\end{equation}
For $A<-u$ the second term of (\ref{sc1}) is indeed smaller,
$\exp(-u^2/2A)<\exp(u)\ll \exp(u+A/2)$. The range where $\sigma_c$
is finite, i.e. the charge density is finite and behaves as a
plasma is where the exponent in the solution is positive (both
$\sigma_c$ and $y$ being small),
\begin{equation}\label{sc3}
\sigma_c\sim y^{\frac{2}{2-K+K^2\sigma}}  \qquad  K-K^2\sigma
-2>0, \qquad \sigma <2/K^2
\end{equation}
and the critical line where $\sigma_c$ vanishes
is $K-K^2\sigma -2=0$ (Fig. 2); the
condition $A<-u$ becomes $\sigma <2/K^2$ (see below). This is
the first, or high temperature regime. In that regime
a standard small fugacity expansion works, the effects
of the width of $\Phi(v)$ are unimportant,
both at the transition and in the disordered phase.

Considering now the second, or low temperature,
regime $A>-u$. Then the first term of (\ref{sc1}) is dominated by the upper
limit, hence both terms of (\ref{sc1}) yield
\begin{equation}\label{sc4}
\sigma_c\sim e^{-u^2/2 A}\sim
y^{1/4 K\sigma} \sigma_c^{1/8\sigma} \qquad \sigma
>\frac{1}{8}, \qquad  A>-u \,.
\end{equation}
Note that this corresponds to the distribution $\Phi(v)$ being
very broad and then the maximum at $v=-u$ dominates the result. For
the finite charge density phase we have now
\begin{equation}\label{sc5}
\sigma_c\sim y^{\frac{2/K}{8\sigma -1}} \qquad \sigma
>\frac{1}{8}\qquad \sigma >2/K^2
\end{equation}
so that the critical line is $\sigma =\frac{1}{8}$ (Fig. 2); the
condition $A>-u$ becomes $\sigma >2/K^2$ (see below).

\begin{figure}[htb]
\begin{center}
\includegraphics[scale=0.4]{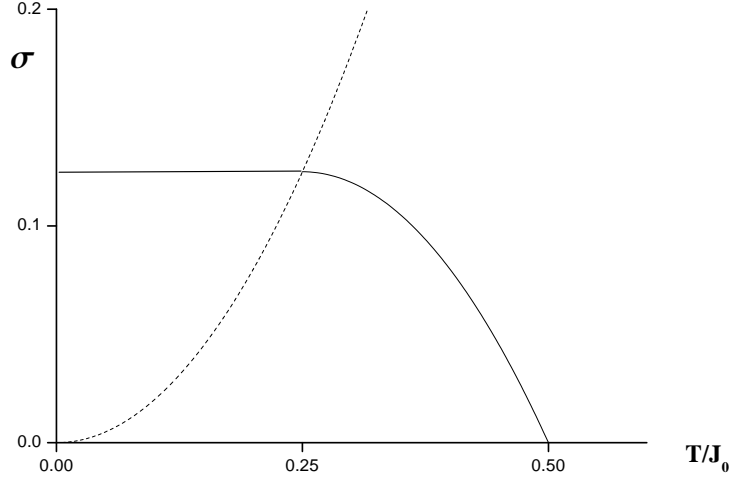}
\end{center}
\caption{Phase diagram for one layer in terms of $\sigma$ and
$T/J_0=1/K$ variables. The full line is the defect transition
given by $K-K^2\sigma-2=0$ at $\frac{1}{4} <1/K<\frac{1}{2}$ and
by $\sigma =\frac{1}{8}$ at $1/K< \frac{1}{4}$. The dashed line
$\sigma =2/K^2$ is a first order transition within the ordered
(low $T$) phase while a crossover line in the disordered phase. }
\end{figure}

The boundary between the regimes (\ref{sc3}) and (\ref{sc5}) is
$A=-u$, which for $\sigma_c \to 0$ is $\sigma =1/4K$, i.e.
$\sigma=\frac{1}{8}$, $K=2$ on the critical line. The form
(\ref{sc3}) is then valid at high temperatures $K<2$ and a sum on
single replica, single charge excitations is sufficient. In the
low temperature regime ($K>2$), where Eq. (\ref{sc5}) is valid,
the summation on all charges in all replicas $n_a = (0,\pm1)$ is
essential in obtaining the correct result. It corresponds to the
physics of the freezing, or prefered localisation of the charges
in deep minima of the random potential.

It is instructive to evaluate the boundary between the regimes
(\ref{sc3}) and (\ref{sc5}) for arbitrary small bare fugacity
$y\ll 1$ also away from the critical line. The non-analytic
behavior of the integral in Eq. (\ref{sc1}) is related to the
divergence of $u$, i.e. it exists in the ordered phase, while
becomes a cross-over line in the disordered phase; this is further
discussed below. Consider then a finite $\sigma_c$ and define
$\sigma_c\sim y^{\gamma}$. For $y\ll 1$ the condition $A=-u$
becomes $\sigma=(1/2K)+(1/\gamma K^2)$. For $A<-u$ we have from
Eq. (\ref{sc3}) $\gamma=2/(2-K+\sigma K^2)$, hence the boundary is
$\sigma=2/K^2$. Similarly, for $A>-u$, using $u=(K\gamma/2+1)\ln
y$ yields $\gamma=(2/K)/(\sqrt{8\sigma}-1)$ and again the boundary
is at $\sigma=2/K^2$. Hence there is a unique boundary between the
two regimes [as included in the conditions for Eqs. (\ref{sc3})
and (\ref{sc5})] which intersects the critical line at
$\sigma=\frac{1}{8}$, $K=2$. The sharpness of this boundary, as
mentioned above, depends on $\sigma_c \to 0$, hence in the
disordered phase it depends on the smallness of $y$, i.e. it is a
crossover line where the charge densities $\sim \sigma_c$ change
from Eq. (\ref{sc3}) to Eq. (\ref{sc5}), a crossover whose width
shrinks with $y$. In the ordered phase $\sigma_c=0$ and formally
the boundary is sharp, although the relevant observable, i.e. the
density, vanishes. One may still observe this transition by a
finite size effect where the $q \to 0$ singularity is cutoff by
the inverse area $1/L^2$ instead of $\sigma_c$, i.e. $
\sigma_c\sim (1/L)^{K-\sigma K^2}$ or $\sim (1/L)^{1/4\sigma}$ in
the two regimes, respectively. This transition is termed as a
freezing transition; it is related to the single directed polymer
transition on a Cayley tree \cite{ct}, to a dynamic transition
\cite{dcpld} and also to a phase transition in a random gauge
Dirac system \cite{rdirac}.

Consider next $\sigma_0$, Eq. (\ref{s0}). The integral is again
dominated by large $|v|$, hence
\begin{equation}\label{s01}
\sigma_0 \approx
\int_0^{\infty}\frac{e^{2u+2v}-v^2/2A}{(1+e^{u+v})^2}\frac{dv}{\sqrt{2\pi
A}}\approx \int_0^{-u}e^{2u+2v-v^2/2A}\frac{dv}{\sqrt{2\pi A}}+
\int_{-u}^{\infty}e^{-v^2/2A}\frac{dv}{\sqrt{2\pi A}}\,.
\end{equation}
The second term is $\sim \exp(-u^2/2A)$ while the first term has a
saddle point at $v=2A$ which is inside the integration range if
$2A<-u$, and then
\begin{equation}\label{s02}
\sigma_0\sim y^2e^{2u+2A}\sim y^2 (\sigma_c)^{K-2\sigma K^2}
\qquad  \sigma <2/3K^2 \,.
\end{equation}
$2A<-u$ implies $\sigma<(1/4K)+1/(2\beta K^2)$ and since also
$A<-u$ we can use $\beta=2/(2-K+\sigma K^2)$, hence $\sigma
<2/3K^2$. Note that $\sigma_0\sim \sigma_c^{2-\sigma K^2} \ll
\sigma_c$ when $\sigma <2/3K^2$, so that $\sigma_0/ \sigma_c$ in
Eq. (\ref{A}) can be neglected. Consider next $2A>-u$ where the
integrals for $\sigma_0$ are dominated by the end points $v=-u$.
The range $-u<2A<-2u$ which corresponds to $2/3K^2<\sigma<2/K^2$
yields
\begin{equation}\label{s03}
  \sigma_0\sim (\sigma_c)^{(1+\sigma K^2/2)^2/(2\sigma K^2)}
  \qquad 2/3K^2<\sigma<2/K^2
\end{equation}
for which again $\sigma_0\ll \sigma_c$ while at $\sigma >2/K^2$ we
have $\sigma_0\sim \sigma_c$. At $\sigma=2/3K^2$ the functional
form of $\sigma_0$ changes, but since near this line $\sigma_0\ll
\sigma_c$ there is no observable singularity.

To conclude, comparison with RG studies \cite{scheidl97,dcpld} shows that the
present variational method, which accounts for
broad fugacity distributions, gives a remarkably accurate
description of the transition and in particular of the freezing phenomena
at low temperature in the single layer model. This is presumably because the
screening effects (neglected in the variational approach)
was shown, via higher order RG, to be very small at low temperature.
In addition it provides a description of the disordered phase.
The RG methods can be extended to many layers but following the joint
distribution of the large set of random fugacities becomes
rapidly difficult as $M$ increases. We thus now turn to the
extension of the variational method to several layers.

\section{variational method - many layers}

\subsection{general case}

We study now the full many-layer system, Eq. (\ref{SG}). We
develop a variational method for $M$ coupled layers which allows
for fugacity distributions, an essential feature in the one-layer
problem. It is explicitly worked out for two layers, describing
the various rod transitions as found by the energy rationale in
section III, as well as a narrow transition region.

We note in particular the form of the interaction term $\exp{i
{\bf n} \cdot {\mbox{\boldmath $\chi$}}({\bf r})}$; the naive
approach would be to restrict to charges ${\bf n}$ with a single
non zero entry, leading to a uniform fugacity term $ - y
\sum_{{\bf r},n,a} \cos({\mbox{\boldmath $\chi$}_{na}}({\bf r}))$
and a diagonal $k$-independent replica mass term. Instead we keep
{\it all} composite charges ${\bf n}$, which allow for variational
solutions with off diagonal and $k$-dependent replica mass terms.
This corresponds respectively to fluctuations of fugacity and $N
>1$ charge rods being generated and becoming relevant.

 We note first that a rod solution is readily obtained from Eq.
 (\ref{Hm}), i.e. we look for N correlated charge on nearest
 layers so that
\begin{equation}
  |n_a({\bf q},k)|^2=|n_a({\bf q}) d \sum_{l=0}^{N-1}e^{ikdl}|^2=
  d^2 |n_a({\bf q})|^2\phi_N(k)
\end{equation}
where $\phi_N(k)$ was defined in (\ref{phiN}). With this
replacement Eq. (\ref{Hm}) has the form of a one-layer system Eq.
(\ref{2d}) with effective parameters
\begin{eqnarray}
&& K_{eff} = \int_k g(k)\phi_N(k) \\
&& \sigma_{eff} =  \frac{\int_k \beta^2 \Delta(k)\phi_N(k)}{
(\int_k g(k)\phi_N(k))^2}\label{seffN}
\end{eqnarray}
The system than has the same phase diagram as for one layer (Fig.
2) with these effective parameters. In particular the $T=0$
transition is at $\sigma_{eff}=\frac{1}{8}$, in agreement with Eq.
(\ref{Nrod}).

 We proceed with the variational scheme and define an optimal Gaussian
 Hamiltonian to approximate Eq. (\ref{SG}) as:
\begin{eqnarray}
&&{\cal H} _0=\frac{1}{2} \int_k \int_{\bf q} \chi_a({\bf
q},k)G_{ab}({\bf q},k)\chi^*_b({\bf q},k) \\
&& G^{-1}_{ab}({\bf q},k))=(G_0)^{-1}_{ab}({\bf q},k) +\sigma
_c(k)\delta _{ab}+\sigma_0(k)
\end{eqnarray}
i.e the self energy can now depend on $k$.

The variational free energy is ${\cal F}_{var}={\cal F} _0+\langle
{\cal H}_{SG}-{\cal H}_0\rangle_0$ where $\langle...\rangle$ is an
average with respect to ${\cal H} _0$ and ${\cal F}_0$ is its free
energy $\beta {\cal F}_0=-\frac{1}{2} Tr \ln G_{ab}({\bf q},k)$.
The Gaussian average has the form
\begin{eqnarray}
F[{\bf n}]\equiv \langle\exp{i{\bf n} \cdot {\mbox{\boldmath
$\chi$}}({\bf r})}\rangle_0 = \exp \{-\frac{1}{2}  \int_k \sum_{a}
|\sum_l n_{l,a} e^{i k d l}|^2 G_c(k) + \frac{1}{2} \int_k
|\sum_{l,a} n_{l,a}e^{ik d l}|^2 A(k)\}
\end{eqnarray}
where we recall that $ \int_k \equiv \frac{1}{M d} \sum_k$ and
$\int_q G_{ab}(q,k)=G_c(k)\delta_{ab}-A(k)$, with
\begin{eqnarray}
&& G_c(k) = \int_q G_c(q,k) = g(k)\ln[\Lambda^2/(4\pi
g(k)\sigma_c(k))]
\nonumber \\
&& A(k) = \int_q G(q,k) = \beta^2 \Delta(k) \ln[\Lambda^2/(4\pi
g(k)\sigma_c(k))]
 - \beta^2 \Delta(k) + g(k)\sigma_0(k)/\sigma_c(k)\,.
 \end{eqnarray}
Extremization of $F_{var}$ yields the  saddle point equation:
\begin{eqnarray}
(\sigma_c)_{ll'} \delta_{ab}+ (\sigma_0)_{ll'} = \xi^{-2} d^{-2} \sum_{{\bf n}}
n_{al} n_{b l'} Y[{\bf n}]F[{\bf n}]
\end{eqnarray}
since the dependence is on $l-l'$, a corresponding Fourier
transform yields
\begin{eqnarray}
\sigma_c(k) \delta_{ab} + \sigma_0(k)= \xi^{-2} d^{-1} \sum_{{\bf n}}
\sum_{l-l'} n_{al} n_{bl'} e^{i k d (l-l')} Y[{\bf n}] F[{\bf n}]
\end{eqnarray}

We can now define $s_a(k) = d \sum_l n_{l,a} e^{i k d l}$.
The $A(k)$ term can be written as an average over
gaussian distributions of fugacities:
\begin{eqnarray}
&& \prod_k \exp{ ( \frac{1}{2 M d^3} |s_a(k)|^2 A(k) ) } = \int
\prod_k \frac{d^2 w_k}{2 \pi} \exp{ ( \frac{1}{M d^2} \sum_k \Re
[\omega_k s_a^*(k)]
- \frac{1}{2 M A(k) d} |\omega_k|^2 ) } \\
&& = \prod_k \langle \exp{ (\frac{1}{M d^2}  \Re [\omega_k
s_a^*(k)] )} \rangle_ {\omega}
\end{eqnarray}
This form allows to
decouple $\sum_{{\bf s}} F[{\bf s}] = \langle Z^m\rangle_{\omega}$ with
\begin{equation}\label{Z4}
  Z=\sum_{\{s_n=0,\pm 1\}} \exp(- \frac{1}{2 d^2} \int_k
\tilde{G}_c(k) |s(k)|^2 + \frac{1}{d} \int_k \Re [\omega_{k} s^*(k) ])
\end{equation}
The variational equations for $m\rightarrow 0$ become
\begin{equation}
\sigma_c(k)= \xi^{-2} d \langle \frac{\partial^2 \ln
Z}{\partial \omega_{k}
\partial \omega^*_{k}} \rangle_{\omega} \,; {\mbox {\hspace{2mm}}}
\sigma_0(k)= \xi^{-2} d \langle |\frac{\partial \ln
Z}{\partial \omega_{k} }|^2 \rangle_{\omega} \label{consistency}
\end{equation}
We will not attempt to solve the general case but rather
present a solution for $M=2$.

\subsection{detailed solution for two layers}

We consider now two layers with uncorrelated and equal disorder on
each layer. The partition sum depends now on the number of $+$ and
$-$ charges on each layer, i.e. on the 8 numbers
$n_{\alpha,\beta}$ where $\alpha, \beta=\pm 1,0$, excluding
$n_{00}$. For the vectors ${\bf n}_1, {\bf n}_2$ in replica space
for each layer, their number for a given collection of
$n_{\alpha,\beta}$ is the combinatorial factor in the following
sum
\begin{eqnarray}
\sum_{{\bf n}_1, {\bf n}_2}Y[{\bf n}_1, {\bf n}_2] F[ {\bf n}_1,
{\bf
n}_2]=&&\sum_{n_{\alpha,\beta}}\frac{m!}{n_{00}!\,\,n_{+0}!...n_{++}!}\exp
[- \beta
E_c\sum_a(n_{a1}^2+n_{a2}^2)-\frac{1}{4d}G_c(0)\sum_a(n_{a1}+n_{a2})
^2\nonumber\\
&&-\frac{1}{4d}G_c(\pi)\sum_a(n_{a1}-n_{a2})^2+\frac{1}{2}A_1(\sum_an_{a1}+n_{a2})^2
+\frac{1}{2}A_2(\sum_an_{a1}-n_{a2})^2]
\end{eqnarray}
where $A_1=A(0)/2d$, $A_2=A(\pi)/2d$ and the sum is restricted to
$\sum_{\alpha,\beta}n_{\alpha,\beta}=m$ . We need then two
fugacity distributions,
\begin{equation}\label{wi}
  \exp[\frac{1}{2}A_i(\sum_an_{a1}\pm n_{a2})^2]=\int
  e^{\omega_1\sum_a(n_{a1}\pm
  n_{a2})}e^{-\omega_i^2/2A_i}\frac{d\omega_i}{\sqrt{2\pi A_i}}
\end{equation}
where the upper (lower) signs corresponds to $i=1$ or $i=2$,
respectively. The sum over $n_{\alpha,\beta}$ has the form of a
"ninomial" expansion, i.e. a power of 9 terms,
\begin{equation}
\sum_{{\bf n}_1, {\bf n}_2}Y[{\bf n}_1, {\bf n}_2]F[ {\bf n}_1, {\bf
n}_2]=\langle Z^m\rangle_{\omega}
\end{equation}
where the average is on both $\omega_1$, $\omega_2$. In terms of
$u_1=-\frac{1}{2} \beta E_c-\frac{1}{4d}G_c(0)$ and
$u_2=-\frac{1}{2} \beta E_c-\frac{1}{4d}G_c(\pi)$ we have
\begin{eqnarray}
  Z=&& 1+e^{u_1+u_2+\omega_1+\omega_2}+e^{u_1+u_2+\omega_1-\omega_2}
  +e^{u_1+u_2-\omega_1+\omega_2}+e^{u_1+u_2-\omega_1-\omega_2}\nonumber\\
  &&+e^{4u_1+2\omega_1}+e^{4u_2+2\omega_2}+e^{4u_1-2\omega_1}
  +e^{4u_2-2\omega_2}
\end{eqnarray}
The equations for the (dimensionless) self mass terms
$\sigma_{c1}=\frac{\xi^2 d}{2}\sigma_c(0)$,
$\sigma_{c2}=\frac{\xi^2 d}{2}\sigma_c(\pi)$ and similarly for
$\sigma_{0i}$ are
\begin{eqnarray}
&&\sigma_{ci}=<\frac{\partial^2 \ln Z}{\partial
\omega_i^2}>_{\omega}\\
&&\sigma_{0i}=<\frac{(\partial Z/\partial
\omega_i)^2}{Z^2}>_{\omega} \,.
\end{eqnarray}
These self masses correspond to length scales, i.e. $\sigma_{c1}^{-1/2}$
is the typical distance between ($++$) charge rods (i.e. a $+$ in layer 2
is on top of a $+$ in layer 1),
while $\sigma_{c2}^{-1/2}$ is the typical distance between ($+-$) charge rods.
In general $\sigma_{c2}\sim [\sigma_{c1}]^{\alpha}$ so that $\alpha=0$
 corresponds to $\sigma_{c1}=0$ with N=2 ($+-$) rod defects, $\alpha=\infty$
 corresponds to $\sigma_{c2}=0$ with N=2  ($++$) rod defects,
 $\alpha=1$ corresponds to the two length scales being equal hence an N=1
state, while other values of $\alpha$ imply the presence of two independent
length scales.

A $N=2$ rod solution is readily obtained by $\sigma_{c2}=0$ so
that $u_2 \to \infty$ and
$Z=1+e^{4u_1+2\omega_1}+e^{4u_1-2\omega_1}$. This is equivalent to
the one layer system with
\begin{eqnarray}
&&K_{eff}=2(K_0\pm K_1)\nonumber\\
&&\sigma_{eff}=\frac{\sigma K_0^2}{2(K_0\pm K_1)^2}
\end{eqnarray}
where the lower sign corresponds to the $+,-$ rod solution
$\sigma_{c1}=0$.

Consider now a general solution of the form $\sigma_{c2}\sim
[\sigma_{c1}]^{\alpha}$ so that near criticality
\begin{eqnarray}
&& u_1 \to -\frac{1}{4}(K_0+K_1)\ln (\Lambda^2/\sigma_{c1}) \qquad
A_1\to \frac{1}{2}\sigma K_0^2\ln (\Lambda^2/\sigma_{c1})\nonumber\\
&&u_2 \to -\frac{1}{4}\alpha(K_0-K_1)\ln (\Lambda^2/\sigma_{c1})
\qquad A_2 \to \frac{1}{2}\alpha \sigma K_0^2\ln
(\Lambda^2/\sigma_{c1})
\end{eqnarray}
Near criticality the $\omega$ integrals are dominated by large
values so that positive and negative integration ranges are
equivalent; furthermore, the $\omega_1,\omega_2>0$ integral is
dominated by exponents where $\omega_1,\omega_2$ appear with
positive sign,
\begin{equation}\label{sd1}
  \sigma_{c1}=\langle \frac{\partial}{\partial
  \omega_1}\frac{\partial Z/\partial
  \omega_1}{Z}\rangle_{\omega}\approx 4\langle \frac{\partial}
{\partial\omega_1}\frac{e^{u_1+u_2+\omega_1+\omega_2}+2e^{4u_1+2\omega_1}}
  {1+e^{u_1+u_2+\omega_1+\omega_2}+e^{4u_1+2\omega_1}+e^{4u_2+2\omega_2}}\rangle
  _{\omega_1,\omega_2>0}
\end{equation}
We focus here on the low temperature behavior where $K_i \to
\infty$ and the integrals are dominated by the maxima of the above
$\partial /\partial \omega_1$. The fraction in Eq. (\ref{sd1}) has
values $0,1,2$ as indicated in Fig. 3 with boundaries shown by the
full lines, assuming for now $u_2>u_1$ (the solution for $u_2<u_1$
can be inferred by the symmetry of the phase diagram under
$K_0,K_1,\alpha \rightarrow K_1,-K_0,1/\alpha$). At the full lines
in Fig. 3 $\partial /\partial \omega_1$ is maximal and dominate
the integral at low temperatures since the Gaussian averaging
factors are very flat. More precisely, we have separated the
$\omega_1$ integral into ranges left and right of these lines and
check in each range that it has no saddle point and is therefore
dominated by $\omega_1$ at the line position. Thus for
$\omega_2<-2u_2$ the integral is dominated by
$\omega_1=-\omega_2-u_1-u_2$ leading to a contribution
\begin{equation}\label{sd2}
  \sigma_{c1}^{(1)}\sim \int_0^{-2u_2}d\omega_2
  e^{-(\omega_2+u_1+u_2)^2/2A_1-\omega_2^2/2A_2}
\end{equation}
while for $\omega_2>-2u_2$ the intgeral is dominated by
$\omega_1=\omega_2+3u_2-u_1$ with the contribution
\begin{equation}\label{sd3}
\sigma_{c1}^{(2)}\sim \int_{-2u_2}d\omega_2
  e^{-(\omega_2+3u_2-u_1)^2/2A_1-\omega_2^2/2A_2}\,.
\end{equation}
The saddle point of this integral is below $-2u_2$, hence it is
dominated by $\omega_2=-2u_2$, i.e. is the same as
$\sigma_{c1}^{(1)}$ if the latter is also dominated by
$\omega_2=-2u_2$, or less than $\sigma_{c1}^{(1)}$ if the latter
has a saddle point within the integration range. Hence
$\sigma_{c1}^{(1)}$ determines the result with
\begin{eqnarray}\label{sd4}
&&\sigma_{c1}\sim e^{-(u_1+u_2)^2/(2A_1+2A_2)} \qquad
(u_2-u_1)A_2<-2u_2A_1 \nonumber\\
&& \sigma_{c1}\sim e^{-(u_1-u_2)^2/(2A_1)-2u_2^2/A_2} \qquad
(u_2-u_1)A_2>-2u_2A_1
\end{eqnarray}

\begin{figure}[t]
\begin{center}
\includegraphics[scale=1.2]{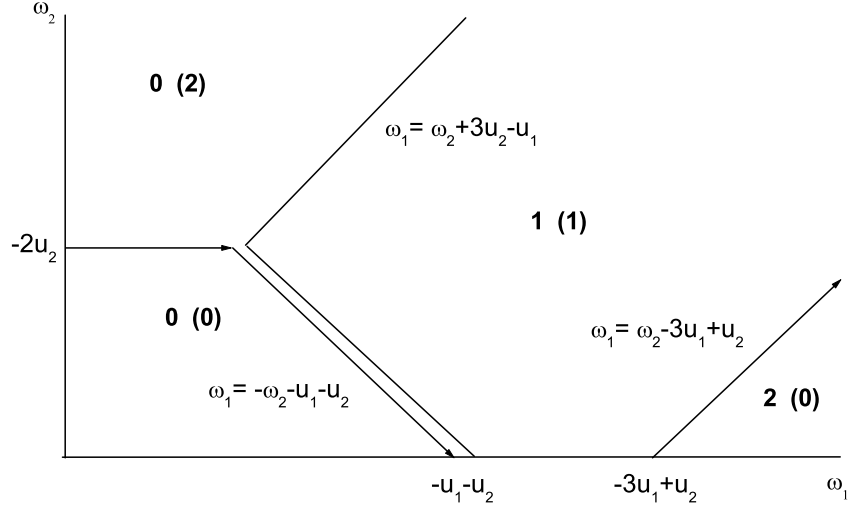}
\end{center}
\caption{Integration ranges for $\sigma_{c1}$ and $\sigma_{c2}$
when $u_2>u_1$ (i.e. $|u_1|>|u_2|$). For $\sigma_{c1}$ the numbers
indicate the fraction value in Eq. (\ref{sd1}) and the full lines
are where the $\omega_1$ integral is dominant. For $\sigma_{c2}$
the numbers in parenthesis indicate the fraction value in Eq.
(\ref{sd5}) and the arrowed lines are where the $\omega_2$
integral is dominant. } \label{Fig3}
\end{figure}

Consider next $\sigma_{c2}$ which for $\omega_1,\omega_2>0$ is
dominated by
\begin{equation}\label{sd5}
  \sigma_{c2}=\langle \frac{\partial}{\partial
  \omega_2}\frac{\partial Z/\partial
  \omega_2}{Z}\rangle_{\omega}\approx 4\langle \frac{\partial}
{\partial\omega_2}\frac{e^{u_1+u_2+\omega_1+\omega_2}+2e^{4u_2+2\omega_2}}
  {1+e^{u_1+u_2+\omega_1+\omega_2}+e^{4u_1+2\omega_1}+e^{4u_2+2\omega_2}}\rangle
  _{\omega_1,\omega_2>0}\,.
\end{equation}
The fraction above has values $0,1,2$ is indicated in Fig. 3 with
boundaries shown by the arrowed lines; at these lines $\partial
/\partial \omega_2$ is maximal and the corresponding $\omega_2$
dominate the integral. Hence for $\omega_1<u_2-u_1$ the integral
is dominated by $\omega_2=-2u_2$ leading to a contribution
\begin{equation}\label{sd6}
  \sigma_{c2}^{(1)}\sim
  \int_0^{u_2-u_1}d\omega_1 e^{-\omega_1^2/2A_1}
  e^{-2u_2^2/A_2}\sim e^{-2u_2^2/A_2}\,.
\end{equation}
The next range is $u_2-u_1<\omega_1<-u_2-u_1$ where
$\omega_2=-\omega_1-u_1-u_2$ dominates, contribution
\begin{equation}\label{sd7}
\sigma_{c2}^{(2)}\sim
  \int_{u_2-u_1}^{-u_2-u_1} d\omega_1 e^{-(\omega_1+u_1+u_2)^2/2A_2}
  e^{-\omega_1^2/2A_1}\,.
\end{equation}
This has a maximum within integration range if
$(u_2-u_1)A_2<-2u_2A_1$ with the result
\begin{equation}\label{sd8}
\sigma_{c2}^{(2)}\sim e^{-(u_1+u_2)^2/(2A_1+2A_2)} \qquad
(u_2-u_1)A_2<-2u_2A_1
\end{equation}
while if $(u_2-u_1)A>-2u_2A_1$ the integral is dominated by its
lower limit $u_2-u_1$ which is then always smaller then
$\sigma_{c2}^{(2)}$. In the range $-u_1-u_2<\omega_1<-3u_1+u_2$
the line of maximum $\omega_2=\omega_1-3u_2+u_1$ is at large
values of $\omega_2$ (see Fig. 3) so should give a small
contribution; in fact integrating this line even from $u_2-u_1$
yields
\begin{equation}\label{sd9}
  \sigma_{c2}^{(3)}\sim
  \int_{u_2-u_1} d\omega_1 e^{-(\omega_1+u_1-3u_2)^2/2A_2}
  e^{-\omega_1^2/2A_1}\sim e^{-2u_2^2/A_2-(u_2-u_1)^2/2A_1}
\end{equation}
where the integrand has a maximum below the integration range and
is therefore dominated by the lower integration limit. The result
in Eq. (\ref{sd9}) equals also to that of the integrand in
(\ref{sd7}) at its lower limit, hence
$\sigma_{c2}^{(3)}\leq\sigma_{c2}^{(2)}$. Finally, the range
$-3u_1+u_2<\omega_1$ has the line of maximum at
$\omega_2=\omega_1+3u_1-u_2$, hence
\begin{equation}\label{sd10}
  \sigma_{c2}^{(4)}\sim
  \int_{-3u_1+u_2} d\omega_1 e^{-(\omega_1+3u_1-u_2)^2/2A_2}
  e^{-\omega_1^2/2A_1}\sim e^{-(3u_1-u_2)^2/A_1}
\end{equation}
where again the integral is dominated by its lower limit. This
result is smaller then Eq. (\ref{sd8}) (it is smaller then the
integrand of (\ref{sd7}) at $\omega_1=-u_1-u_2$, hence the latter
is bigger if it has a maximum within integration range).

\begin{figure}[b]
\begin{center}
\includegraphics[scale=0.3]{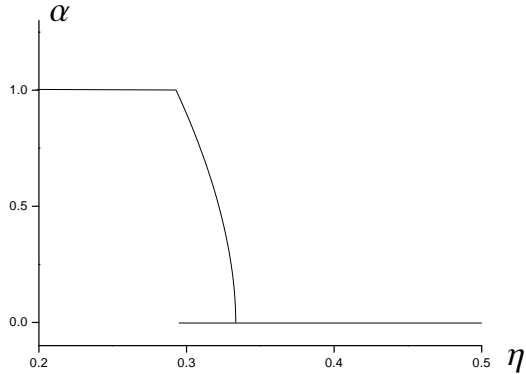}
\end{center}
\caption{Two layer solutions for the exponent in $\sigma_{c2}\sim
[\sigma_{c1}]^{\alpha}$ in terms of the anisotropy $\eta$. In the
range $1-1/\sqrt{2}<\eta<1/3$ two solutions coexist.}
\end{figure}

\begin{figure}[t]
\begin{center}
\includegraphics[scale=0.3]{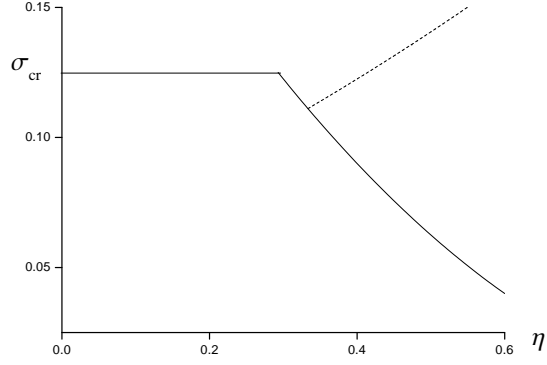}
\end{center}
\caption{The critical disorder $\sigma_{cr}$ for a two layer
system. At $\eta<1-1/\sqrt{2}$ the transition is to a $N=1$ phase
at $\sigma_{cr}=\frac{1}{8}$. For $\eta>1-1/\sqrt{2}$ at
$\sigma_{cr}=\frac{1}{4}(1-\eta)^2$ the transition is either to an
$N=2$ rod phase at $\eta>1/3$ or, for $1-1/\sqrt{2}<\eta<1/3$ a
mixed phase $\sigma_{c2}\sim [\sigma_{c1}]^{\alpha}$ is possible.
At $\eta>1/3$ the $N=2$ rod solution disintegrates into the $N=1$
phase at $\sigma_{cr}^{(1,2)}=(1+\eta)^2/16$.}
\end{figure}

Collecting all terms we have,
\begin{eqnarray}\label{sd11}
&& \sigma_{c2}\sim \max [e^{-2u_2^2},\,
e^{-(u_1+u_2)^2/(2A_1+2A_2)}] \qquad  (u_2-u_1)A_2<-2u_2A_1
\nonumber\\
&& \sigma_{c2}\sim \max [e^{-2u_2^2},\,e^{-(3u_1-u_2)^2/A_1}]
\qquad  (u_2-u_1)A_2>-2u_2A_1 \,.
\end{eqnarray}
Eqs. (\ref{sd4},\ref{sd11}) can be written in terms of $\alpha$
and an anisotropy parameter $\eta=K_1/K_0>0$ (For $\eta<0$ we note
that the solutions are symmetric under $\eta, \alpha \rightarrow
-\eta, 1/\alpha$). For $\eta<\frac{1+\alpha}{3+\alpha}$
\begin{eqnarray}\label{sc11}
\sigma_{c1}&&\sim [\sigma_{c1}]^{\frac{(1+\alpha+\eta-\eta
\alpha)^2}{16(1+\alpha)\sigma}}\nonumber\\
\sigma_{c2}&&\sim \max \{\,[\sigma_{c1}]^{\frac{\alpha
(1-\eta)^2}{4\sigma}}, \,\, [\sigma_{c1}]^{\frac{(1+\alpha +\eta
-\eta \alpha)^2}{16(1+\alpha)\sigma}}\}
\end{eqnarray}
while for $\eta>\frac{1+\alpha}{3+\alpha}$ we have
\begin{eqnarray}\label{sc12}
\sigma_{c1}&&\sim [\sigma_{c1}]^{\frac{(1-\alpha +\eta +\eta
\alpha)^2+4\alpha (1-\eta)^2}{16\sigma}}\nonumber\\
\sigma_{c2}&&\sim \max \{\,[\sigma_{c1}]^{\frac{\alpha
(1-\eta)^2}{4\sigma}}\,\, , [\sigma_{c1}]^{\frac{(3+3\eta -\alpha
+\eta \alpha)^2}{16\sigma}}\}\,.
\end{eqnarray}
Some inspection shows that the latter equation has no solutions
(except with $\alpha =0$, see below) while for Eq. (\ref{sc11}) we
have the following solutions (see Fig. 4): (i) The 2nd term of
$\sigma_{c2}$ ($\sim \sigma_{c1}^{\alpha}$) identifies
$\sigma_{c1}$ exponents and leads to $\alpha =1$ and criticality
at $\sigma_{cr}=\frac{1}{8}$, i.e. the independent layer solution
$N=1$. The 2nd term of the $\sigma_{c2}$ line is the maximal one
if $\eta <1-1/\sqrt{2}$. (ii) The 1st term of $\sigma_{c2}$
identifies exponents as $(1-\eta)^2/4=(1+\alpha+\eta-\eta
\alpha)^2/16(1+\alpha)$. This term dominates in the $\sigma_{c2}$
line if $\alpha <1$, hence the solution
\begin{equation}\label{eta}
\eta= \frac{2-\sqrt{1+\alpha}}{2+\frac{1-\alpha}{\sqrt{1+\alpha}}}
\end{equation}
exists for $1-1/\sqrt{2}<\eta<1/3$ with
$\sigma_{cr}=\frac{1}{4}(1-\eta)^2$. (iii) Finally $\alpha=0$ is
possible, i.e. $\sigma_{c1}\equiv 0$ and an onset of just the
$k=\pi$ component $\sigma_{c2}$. The solution is then of charges
correlated between layers, i.e. the $N=2$ rod phase. Criticality
is at $\sigma_{cr}=\frac{1}{4}(1-\eta)^2$, and from both Eqs.
(\ref{sc11},\ref{sc12}) this solution is valid at all $\eta$
provided it precedes the solution (i) with
$\sigma_{cr}<\frac{1}{8}$, hence $\eta
>1-1/\sqrt{2}$.

The solutions (i) and (iii) reproduce the energy rationale. We
have found here an additional solution (ii) with a nontrivial new
exponent $\alpha$ in a narrow range $1-1/\sqrt{2}<\eta<1/3$. This
solution is a continuous interpolation in $\alpha$ between the
$N=1$ solution ($\alpha =1$ at $\eta <1-1/\sqrt{2}$) and the $N=2$
rod solution ($\alpha =0$ at $\eta >1/3$). Both solutions (ii) and
(iii) have the same $\sigma_{cr}$, hence they may be degenerate.

Solution (iii) allows for an additional phase transition
corresponding to the onset of $\sigma_{c1}$, i.e. the $N=2$ rods
decompose into independent $N=1$ charges on each layer. When
$\sigma_{c2}\neq 0 $, $u_2$ and $ A_2$ are finite, hence the
divergent terms in the exponent of Eq. (\ref{sd4}) yield
$\sigma_{c1}\sim e^{-u_1^2/2A_1}\sim
\sigma_{c1}^{(1+\eta)^2/16\sigma}$, hence
$\sigma_{cr}^{(1,2)}=(1+\eta)^2/16$ allows the onset of
$\sigma_{c1}$ at $\eta>1/3$ (dashed line in Fig. 5). The energy
rationale gives a somewhat higher
$\sigma_{cr}^{(1,2)}=\eta_1^2/[4(\sqrt{2}-1)^2]$ for this $N=2$ to
$N=1$ transition.

\begin{figure}[b]
\begin{center}
\includegraphics[scale=0.5]{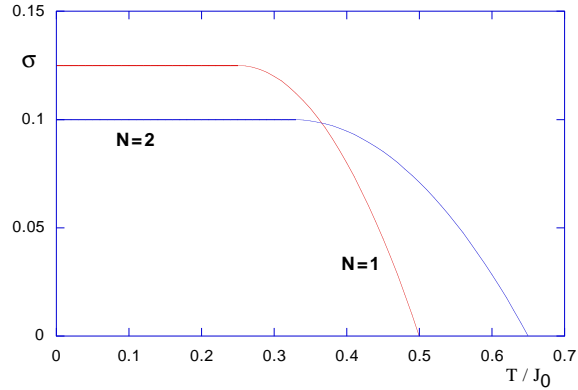}
\end{center}
\caption{ Phase diagram for the onset of the $N=1,2$ instabilities
for anisotropy $\eta=0.35$. At low $T$ two distinct transitions
are possible, the first being to the rod $N=2$ phase. At high $T$
the independent layer $N=1$ transition dominates and eliminates
the $N=2$ phase.}
\end{figure}

Finally we consider the disorder-temperature phase diagram. The
high temperature part of the phase boundary is determined by low
order renormalization group as disorder is well behaved. Thus, in
either Coulomb gas formulation \cite{dcpld} or in sine-Gordon
formulation we find the recursion with scale $\ell$
\begin{equation}\label{RG}
\frac{\partial Y[n]}{\partial
\ell}=Y[n]\{2-\sum_{l,l',a}n_{a,l}n_{a,l'}K_{l-l'}+\sigma
K_0^2\sum_l[\sum_{a}n_{a,l}]^2\}
\end{equation}
The $N=1$ solution is determined by one nonzero entry, hence
$2-K_0+\sigma K_0^2=0$; for $N=2$ the solution corresponds to one
nonzero entry per two layers, with the relative sign $\mp$, hence
$2-2K_0\pm 2K_1+2\sigma K_0^2=0$. For $\eta>0$ the dominant
transition (i.e. the one at lower temperature) has the upper sign,
corresponding to $\sigma_{2c}$ with $k=\pi$. At $\sigma =0$ this
has a critical temperature lower than that of $N=1$ since
$K_0=1/(1-\eta)<2$ for $\eta <\frac{1}{2}$. Therefore the range of
low $\sigma$ is dominated by the usual $N=1$ transition. In Fig. 6
we demonstrate the phase diagram with $\eta=0.35$ where the phases
$N=1,2$ compete.


\section{Application to superconductors}

\subsection{Layered superconductor without disorder}
The standard model for layered superconductors is the Lawrence
Doniach model in terms of the superconducting phases on each layer
and the electromagnetic vector potential. The latter can be
integrated out \cite{H3} leading to an effective Hamiltonian in
terms of pancake vortices, i.e. point singularities in each layer,
and a nonsingular Josephson phase. We consider here the case
without Josephson coupling, where the pancake vortices are not
coupled to the Josephson phase. If $n({\bf r},l)$ is an integer
field of $\pm 1, 0$ corresponding to the location of pancake
vortices then the vortex Hamiltonian is \cite{H3}

\begin{equation}\label{Hv}
H_{v} = \frac{1}{2} \sum_{{\bf r} \neq {\bf r}'} \sum_{ll'} n({\bf
r},l)G_v({\bf r}-{\bf r}',l-l') n({\bf r}',l')
+ E_c \sum_{{\bf r},l} n({\bf r},l)^2 \\
\end{equation}
with:
\begin{eqnarray}
&& G_v(q,k) = \frac{\Phi_0^2 d^2}{4 \pi \lambda_{ab}^2} \frac{1}{q^2}
\frac{1}{1 + f(q,k)} \label{Gv}\\
&& f(q,k) = \frac{d}{4 \lambda_{ab}^2 q} \frac{\sinh(qd)}{
\sinh^2(\frac{qd}{2}) + sin^2(\frac{kd}{2})}
\end{eqnarray}
where $\lambda_{ab}$ is the magnetic penetration length parallel
to the layers and $G_v(q,k)=d \sum_l \int d^2 r G_v(r,l) e^{i k d
l + i {\bf q} \cdot  {\bf r}}$. The core energy is estimated as
\cite{hu,olive} $E_c\approx (0.04-0.2)\tau$ where $\tau= \Phi_0^2
d/(4 \pi^2 \lambda_{ab}^2)$.

Note that the $k=0$ mode is screened, i.e. $G_v(q,k)$ is
nonsingular at $q=0$. All the other modes are unscreened and lead
to logarithmic interactions. This is because no screening current
can go along $z$ (in the absence of Josephson coupling) and thus
two pancakes in two different layers cannot screen each others.

In presence of an external field $B$ along $z$ a flux lattice with
a unit cell area $a^2=\Phi_0/B$ is formed. The flux lattice is
composed of pancake vortices, i.e. point singularities, which are
displaced from the $p$-th line position ${\bf R}_p$ at the $l$-th
layer into ${\bf R}_p+{\bf u}_p^l$; its Fourier transform is
\begin{equation}
{\bf u}({\bf q},k) = d a^2 \sum_l \sum_p {\bf u}_p^l e^{i {\bf
q}\cdot {\bf R}_p + i k d l}
\end{equation}
Expanding Eq. (\ref{Hv}) to second order in ${\bf u}_p^l$ yields
the elastic Hamiltonian of the form,
\begin{eqnarray}\label{Hel}
&& H_{el} = \frac{1}{2}  \int_k \int_{{\bf q}} [ D_L(q,k)
|u_L({\bf q},k)|^2 + D_T(q,k) |u_T({\bf q},k)|^2]\,.
\end{eqnarray}
We will be mainly interested in the case of no Josephson coupling,
where the following exact expression holds:
\begin{eqnarray}
&& D_L(q,k) P^L_{\alpha \beta}(q) + D_T(q,k) P^T_{\alpha \beta}(q) \\
&& = \frac{1}{a^4 d^2} ( q_\alpha q_\beta G_v(q,k)
+ \sum_{Q \neq 0} ( (Q+q)_\alpha (Q+q)_\beta G_v(Q+q,k) - Q_\alpha Q_\beta G_v
(Q,0) ))
\end{eqnarray}
provided we add a short distance cutoff in plane, i.e replace
$G_v(q,k) \to G_v(q,k) e^{-q^2 \xi^2/2}$. The conventional elastic
moduli are then identified as:
\begin{eqnarray}
&& D_L(q,k) = q^2 c_{11}(q,k) + k_z^2 c_{44}^L(q,k) \\
&& D_T(q,k) = q^2 c_{66}(q,k) + k_z^2 c_{44}^T(q,k)
\end{eqnarray}
where $k_z^2 = \frac{4}{d^2} sin^2(\frac{kd}{2})$. For zero
Josephson coupling it is found \cite{Goldin}, where leading terms
in $q^2$ are retained,
\begin{eqnarray}
&& c_{66}(q,k) = \frac{B \Phi_0}{(8 \pi \lambda_{ab})^2}  \\
&& c_{11}(q,k) + c_{66}(q,k) = \frac{B^2}{4 \pi} \frac{1}{1 +
\lambda_{ab}^2 (q^2 + k_z^2)} \\
&& c_{44}^{L}(q,k) = \frac{B^2}{4 \pi} \frac{1}{1 +
\lambda_{ab}^2 (q^2 + k_z^2)} + c_{44}^{T}(k) \\
&& c_{44}^{T}(k) = \frac{1}{2} (\frac{1}{d a^2})^2 \frac{1}{k_z^2}
\sum_{{\bf Q} \neq {\bf 0}} (G_v(Q,k) - G_v(Q,0)) Q^2 \approx
\frac{2 B \Phi_0}{(8 \pi \lambda_{ab}^2)^2} \frac{1}{k_z^2}
\ln\frac{1 + k_z^2/Q_0^2}{1 + \xi^2 k_z^2}
\end{eqnarray}
and the last form is in the limit $d \ll a, \lambda_{ab}$. We note
that with Josephson coupling the results for $c_{66},c_{11}$ are
unchanged, while $c_{44}^{L,T}$ are modified with a stronger
effect \cite{Goldin} on $c_{44}^{T}$.

We consider first the defect transition in the pure system
\cite{Dodgson}. This refers to the proliferation of vacancy
interstitial pairs (VI), thereby destroying the superconducting
order parallel to the layers. These defects correspond to
additional pancake vortices, denoted by $s_l({\bf r})$ on top of
the ones forming the flux lattice. These defects couple to the
lattice via the same coupling of Eq. (\ref{Hv}),
\begin{eqnarray}
{\cal H}_{vac}=\sum_{{\bf r},p,l,l'}
s_{l}({\bf r})
G_v({\bf R}_p+{\bf u}_p^{l'} -{\bf r}, l-l')
\end{eqnarray}
To 0-th order in ${\bf u}_l^n$ the defects feel
a periodic potential:
\begin{eqnarray}
{\cal H}_{vac}^{(0)}=\sum_{{\bf r},p,l,l'} s_l({\bf r}) G_v({\bf
R}_p -{\bf r}, l-l ')
\end{eqnarray}
which fixes the defect position in a unit cell, hence fluctuations
of $s({\bf q},k) = d \sum_l \sum_{\bf r} s_l({\bf r}) e^{i k d l +
i {\bf q} \cdot {\bf r}}$ involve only ${\bf q}$ in the first
Brillouin zone (BZ); in the following (and in Eq. \ref{Hel}) all
$q$ integrals are restricted to the first BZ. Note that for
vacancies the periodic potential has minima on the flux lines,
while for interstitials the minima are in the middle of the unit
cell. Hence, the core energies of vacancies and interstitials
differ, but as they come in pairs, $E_c$ refers to an average of
these core energies. For an isolated pancake vortex \cite{hu}
$E_c\approx (0.1-0.2)\tau$, while in presence of a flux lattice with
local relaxation leads to \cite{olive} $E_c\approx 0.04\tau$.

Expanding to first order, one finds with the above definitions:
\begin{eqnarray}
&& {\cal H}_{vac}(s,u) = {\cal H}_{vac}^{(0)}(s) + {\cal H}_{vac}^{(1)}(s,u) +
O(s u^2) \\
&& {\cal H}_{vac}^{(1)}(s,u) = \frac{1}{a^2 d^2} \int_{k} \int_{q}
s({\bf q},k) G_v(q,k) (- i {\bf q}) \cdot {\bf u}(-{\bf
q},-k)\label{Hvac1}
\end{eqnarray}
The total energy is thus:
\begin{eqnarray}
&& H_{el}(u) + {\cal H}_{vac}^{(1)}(s,u) + {\cal H}_{v}(s)
= \frac{1}{2}  \int_k \int_{q} D_T(q,k) |u_T({\bf q},k)|^2 \\
&& + \frac{1}{2}  \int_k \int_{q} [( D_L(q,k) |u_L({\bf q},k)|^2 +
\frac{1}{d^2} s({\bf q},k) G_v(q,k) s(-{\bf q},-k) + \frac{2}{a^2
d^2} s({\bf q},k) G_v(q,k) (- i q) u_L(-{\bf q},-k)]\label{h1}
\end{eqnarray}
One can either minimize it to find the (purely longitudinal)
deformation of the lattice induced by the defect,
\begin{eqnarray}
{\bf u}_{vac}({\bf q},k)=i{\bf q}s({\bf q},k)G_v(q,k)/a^2 d^2 D_L(q,k)
\label{u}
\end{eqnarray}
and compute ${\cal H}_{el} + {\cal H}_{vac}+ {\cal H}_{v}$ at the
minimum or, since it is Gaussian, simply integrate out the
displacements $u_L({\bf q},k)$. One finds that the screening of
the vortices by the longitudinal displacements of the lattice
results in an effective interaction energy between the defects:

\begin{eqnarray}
&& {\cal H}^{eff}_{v}(s) = \frac{1}{2 d^2} \int_k \int_{q}
s({\bf q},k) G^{eff}_v(q,k) s(-{\bf q},-k) \\
&& G^{eff}_v(q,k) = G_v(q,k) (1 - \frac{q^2 G_v(q,k)}{a^4 d^2
D_L(q,k)})\label{Geff}
\end{eqnarray}

One can connect with the notations of the previous sections
($\beta=1/k_BT$):
\begin{eqnarray}
g(k) = \frac{\beta}{4 \pi} \lim_{q \to 0} q^2 G^{eff}_v(q,k)
\end{eqnarray}
The pure defect transition thus occurs when:
\begin{eqnarray}\label{Tdef}
K_{eff} = \int_k g(k) = 2  \quad \leftrightarrow \quad
T_{def} = \frac{1}{8 \pi} \int_k G^{eff}_v(k)
\end{eqnarray}
where we define:
\begin{eqnarray}
&& G_v(k) = \lim_{q \to 0} [q^2 G_v(q,k)] =\frac{\Phi_0^2 d^2}{4
\pi}
\frac{k_z^2}{1 + \lambda_{ab}^2 k_z^2}  \\
&& G^{eff}_v(k) = \lim_{q \to 0} [q^2 G^{eff}_v(q,k)] = G_v(k) (1
- \frac{G_v(k)}{a^4 d^2 D_L(0,k)})
\end{eqnarray}
where we recall $k_z^2 = \frac{4}{d^2} sin^2(\frac{kd}{2})$.

It is instructive to consider the "unscreened defect transition"
temperature, i.e. formation of pancake vortices in the absence of
an external field. This is denoted as the vortex transition
\cite{H3} with the onset temperature at
\begin{eqnarray}\label{Tv}
&& T_v =  \frac{1}{8 \pi} \int_k G_v(k)
= \frac{\Phi_0^2 d}{2 (4 \pi \lambda_{ab})^2} h(d/(2 \lambda_{ab})) \\
&& h(y) = \frac{1}{2 \pi} \int_{-\pi}^{\pi} dx
\frac{1}{1 + \frac{y^2}{sin^2(\frac{x}{2})}} =
1 - \frac{y}{\sqrt{1 + y^2}}
\end{eqnarray}
In particular for $d \ll \lambda$ one has:
\begin{eqnarray}
T_v \approx \frac{\Phi_0^2 d}{2 (4 \pi
\lambda_{ab})^2}=\tau/8 \,.
\end{eqnarray}
The actual superconducting transition is at $T_c$ with
$T_v<T_c<T_f$ where $T_f$ is the fluxon transition temperature,
where Josephson decoupling would occur in the absence of pancake
defects \cite{H3}.

To compare the vortex transition with melting we use a Lindemann
type criterion (with only transverse modes).
\begin{eqnarray}
c_L^2 a^2 = \langle u^2 \rangle =
T_m \int_k \int_{q,BZ} \frac{1}{c_{66} q^2 + c_{44}^T k_z^2}
= \frac{T_m}{4 \pi c_{66}} \int_k
\ln(1 + \frac{4 \pi c_{66}}{c_{44}^T a^2 k_z^2})
\end{eqnarray}
Using a circular BZ of volume $(2\pi/a)^2$, hence $0<q^2 <4
\pi/a^2$,
\begin{eqnarray}\label{Tm}
&& T_m \approx
\frac{4 \pi c_L^2}{A} c_{66} a^2 d  = \frac{4 \pi c_L^2}{A}
\frac{\Phi_0^2 d}{(8 \pi \lambda_{ab})^2} \nonumber \\
&& A = d \int_k \ln(1 + \frac{4 \pi c_{66}}{c_{44}^T a^2 k_z^2}) =
d \int_k \ln(1 + \frac{\phi_0^2}{16 \pi a^4 \lambda^2 D_T(0,k)})
\end{eqnarray}
where in the last equation we have used the dispersionless value
of $c_{66}$ valid for $a\gg d$. The scales of the vortex and
melting transitions are the same, their ratio being $T^0_{def}/T_m
= A/4 \pi c_L^2$. Hence the condition that the defect transition
occurs before melting and can thus be consistently described is
that $G^{eff}_v(q,k) \ll G_v(q,k) $.

Let us now study the true transition with screening. One denotes
$D_{L,T}(k) = D_{L,T}(0, k) = k_z^2 c_{44}^{L,T}(0,k)$
respectively. Using the above result, one finds in the $q \to 0$
limit the exact expressions:
\begin{eqnarray}
&& D_T(k) = \frac{1}{2} (\frac{1}{d a^2})^2
\sum_{{\bf Q} \neq {\bf 0}} (G_v(Q,k) - G_v(Q,0)) Q^2  \\
&& D_L(k) = D_T(k) + \frac{1}{a^4 d^2} G_v(k)
\end{eqnarray}
One thus has:
\begin{eqnarray}
&& G_v^{eff}(k) = G_v(k) \frac{D_T(k)}{D_L(k)}
=  \frac{\Phi_0^2 d^2}{4 \pi} \frac{k_z^2}{1 + \lambda_{ab}^2 k_z^2}
\frac{\epsilon(k)}{1 + \epsilon(k)} \\
&& \epsilon(k) = a^4 d^2 D_T(k)/G_v(k)\,.
\end{eqnarray}
Hence the condition for $T_{def}\ll T_m$, which justifies our
description of the defect transition, is $\epsilon(k)\ll 1$.
We note also that for a single 2D layer there is no tilt modulus
for the FL and $\epsilon=0$; hence a 2D FL has VI's at any finite temperature.

 Let us first consider the regime $2 \pi d \ll a<\lambda$
relevant for layered superconductors. As shown in the Appendix one
has in this regime:
\begin{eqnarray}
D_T(k) \approx \frac{\Phi_0^2}{32 \pi^2 \lambda_{ab}^4 a^2}
\ln\frac{1 + k_z^2/Q_0^2}{1 + {\bar d}^2 k_z^2} + \frac{\Phi_0^2
d^2}{8 Q_0^2 a^4 \lambda_{ab}^4} k_z^2 \theta (d-\xi/2\pi)
\end{eqnarray}
where ${\bar d}=\max(d,\xi/2\pi)$ and for $d< \xi$ only the first
term contributes. This yields:
\begin{eqnarray}
\epsilon(k) \approx  \frac{a^2}{8 \pi \lambda_{ab}^4}
\frac{1+\lambda_{ab}^2 k_z^2 }{k_z^2} \ln(\frac{1 + k_z^2/Q_0^2}{1
+ {\bar d}^2k_z^2}) + \frac{\pi d^2 (1+\lambda_{ab}^2 k_z^2) }{2
Q_0^2 \lambda_{ab}^4}\theta (d-\xi/2\pi) \,.
\end{eqnarray}
Note that the relative contribution of the second term becomes
significant only for $k \sim 1/d$. The condition that $\epsilon(k)
\ll 1$ for all $k$ is thus met for $\lambda/a$ sufficiently large
(high enough field) as:
\begin{eqnarray}
\frac{\lambda^2}{a^2} > \frac{1}{4 \pi} \ln(\frac{2 c}{d Q_0})
\end{eqnarray}
where $c$ is a constant of order $O(1)$ (which can be estimated
from above, with $c=1$ when $d<\xi/2\pi$). As long as $\epsilon(k)
\ll 1$ we find that in all regimes one can estimate:
\begin{eqnarray}\label{Geff1}
&& G_v^{eff}(k) \approx \frac{\Phi_0^2 d^2}{4 \pi} \frac{k_z^2}{1
+ \lambda_{ab}^2 k_z^2} \epsilon(k) \approx \frac{\Phi_0^2 a^2
d^2}{32 \pi^2 \lambda_{ab}^4} \ln(1 + k_z^2/Q_0^2)
\end{eqnarray}
This yields the estimate of the defect transition for $2\pi d \ll
a<\lambda $, using Eq. (\ref{Tdef}) at $2\pi d\ll a$,
\begin{eqnarray}\label{T0}
T_0=T_{def} \approx \frac{\Phi_0^2 d a^2}{128 \pi^3
\lambda_{ab}^4} \ln(a/d) \qquad 1 \ll \frac{a}{d} \ll
\frac{\pi}{c} e^{ 4 \pi \lambda_{ab}^2/a^2} \,.
\end{eqnarray}
We use this $T_0$ as a convenient scale below.
We note that (\ref{Geff1}) is weakly $k$ dependent, hence small
 anisotropy $\eta$ (Fig. 1) and the one-layer N=1 transition
dominates.

We now estimate the defect transition temperature in the other
limit $2 \pi d > a$ relevant for multilayers. As shown in the
Appendix one has then:
\begin{eqnarray}
D_T(k) =  \frac{\Phi_0^2 Q_0 d}{2 \pi a^4} k_z^2 \frac{e^{- Q_0
d}}{(1+ \frac{2 \lambda^2 Q_0}{d})^2}
\end{eqnarray}
 This yields:
\begin{eqnarray}
\epsilon(k) =  Q_0 d e^{- Q_0 d} \frac{1 + \lambda^2 k_z^2}{ 2(1+
\frac{2 \lambda^2 Q_0}{d})^2}
\end{eqnarray}
The condition for $\epsilon(k) \ll 1$ for all $k$ is satisfied
when $2 \pi d \gg a$. Thus in this limit one finds:
\begin{eqnarray}\label{Geff11}
&& G_v^{eff}(k) \approx \frac{\phi_0^2 d^2}{2 \pi} k_z^2 \frac{ d
Q_0 e^{- d Q_0}}{ (1 +\frac{2 \lambda^2}{d^2} Q_0 d)^2 }
\end{eqnarray}
yielding:
\begin{eqnarray}
T_{def} \approx \frac{ \phi_0^2}{8 \pi^2 d} \frac{d Q_0 e^{- d
Q_0}}{ (1 +\frac{2 \lambda^2}{d^2} Q_0 d)^2 }
\end{eqnarray}
using $\int_k k_z^2 = 2/d^3$. If in addition $\lambda > \sqrt{a
d}$ one finds:
\begin{eqnarray}
T_{def} \approx \frac{ \phi_0^2 d^2 a}{64 \pi^3 \lambda^4} e^{- 2
\pi d/a}\,.
\end{eqnarray}
The melting criteria Eq. (\ref{Tm}) has now $c_{44}^T$ which is
exponentially small $\sim e^{-Q_0d}$, however it enters the
logarithm in Eq. (\ref{Tm}), i.e. $A\approx 2\pi d/a$ is large and
$T_m\approx c_L^2\phi_0^2a/(32\pi^2 \lambda_{ab}^2)$, hence
$T_{def}\ll T_m$ for all $2\pi d\gg a$.
We note that Eq. (\ref{Geff11}) implies a significant interlayer
coupling with $\eta$ (see Fig. 1) close to $0.5$; hence disorder
 favors rod phases at low temperatures.

\subsection{model with disorder}

We have seen in section IV that disorder can affect a Coulomb gas
transition if its correlation diverges at least logarithmically
with distance. Therefore, disorder that couples directly to
pancake defects has a finite correlation and has no effect on the
defect transition. In particular the vortex transition Eq.
(\ref{Tv}) in the absence of an external field is disorder
independent. The presence of a flux lattice deformed by point
disorder leads to a significant change in the disorder as seen by
pancake defects. Since each pancake composing the flux lattice is
a charge interacting logarithmically with the pancake defects,
a displaced pancake is equivalent to an addition of $+,-$ charges,
i.e. a dipole. Hence a disorder deformed flux lattice leads to a
quenched dipole disorder seen by the defects, leading to
logarithmically correlated disorder.

We consider first disorder within the finite Larkin scale where
one can expand in displacement, resulting in a random force
$f_l({\bf r})$
\begin{eqnarray}
&& H_{dis} = - \sum_{l} \int d^2{\bf r} {\bf f}_l({\bf r})
u_{L}(l,{\bf r}) \\
&& \overline{f_l({\bf r}) f_{l'}({\bf r}')} = F_{l-l'}
\delta^2({\bf r} - {\bf r}')
\end{eqnarray}
where we display only the coupling to the longitudinal component,
$u_L(l,{\bf r})$ (being a suitable continuation of $u_p^l$) and
$f_l({\bf r})$ is the longitudinal disorder component; only the
longitudinal mode $u_L(l,{\bf r})$ couples to the defects (Eq.
\ref{Hvac1}). One can write it in Fourier components:
\begin{eqnarray}\label{Hdis}
&& H_{dis} = - \frac{1}{d} \int_k \int_q f(k,{\bf q}) u^*(k,{\bf q}) \\
&& \overline{ f(k,{\bf q}) f(k',{\bf q}') } = (2 \pi)^3
\delta^2({\bf q}+{\bf q}') \delta(k+k') F(k)
\end{eqnarray}
where $F(k) = d \sum_l F_{l} e^{i k d l}$; note that for finite
$M$ one has $\int_k \equiv \frac{1}{d M} \sum_k$ and $2 \pi
\delta(k+k') \equiv d M \delta_{k,k'}$. It is useful to relate
$F(k)$ to a previously used \cite{H2,H1} dimensionless disorder
parameter $s$ representing point disorder uncorrelated between
layers. The replicated action has $(32\pi s T_0^2/da^4)\int_{{\bf
q},k}u_a({\bf q},k)u_b^*({\bf q},k)$ with $T_0$ from Eq.
(\ref{T0}). Replicating Eq. (\ref{Hdis}) identifies $F(k)=64\pi s
d T_0^2/a^4$.

We now consider the total energy $H_{el}(u) + H_{vac}(s,u) +
H_{dis}(u)$ and determine the $u$ configuration in presence of
both disorder and defects. The part involving longitudinal
displacements reads:

\begin{eqnarray}
H_{tot} = \int_{{\bf q},k}\frac{1}{2 d^2} s({\bf q},k) G_v( q,k)
s({\bf q},k)^* +&& \frac{1}{d^2 a^2} s({\bf q},k) G_v(q,k) (-iq)
u_L^*({\bf q},k) + \nonumber\\
&&\frac{D_L(q,k)}{2} |u_L({\bf q},k)|^2 - \frac{1}{d} f({\bf q},k)
u_L^*({\bf q},k) \nonumber
\end{eqnarray}
and we neglect the random potential seen by the defect itself
(which is short range). The relaxed phonon field at the minimum
energy is:
\begin{eqnarray}
{\bf u}_{L}({\bf q},k)= i {\bf q} s({\bf q},k) \frac{G_v(q,k)}{a^2
d^2 D_L(q,k)} + \frac{1}{d D_L(q,k)} f({\bf q},k) \label{ud}
\end{eqnarray}
 Computing the energy for the defects at the
minimum (or equivalently integrating out the displacements) yields
the same screened interaction $G_v^{eff}(q,k)$ between defects as
before and in addition yields the coupling of the vacancy to
disorder (through the lattice) as in the starting model which
allows to identify the correlator $\Delta(k)$ introduced in
Section II:

\begin{eqnarray}
&& H_{vdis} = - \sum_{{\bf r},l} V_l({\bf r}) s_l({\bf r})
= - \frac{1}{d} \int_{k,{\bf q}} V({\bf q},k) s^*({\bf q},k) \\
&& V({\bf q},k) = i {\bf q} \cdot f(q,k) \frac{G_v(q,k)}{a^2 d^2 D_L(q,k)} \\
&& \overline{V({\bf q},k) V({\bf q}',k')}=
(2 \pi)^2 \delta({\bf q} + {\bf q}')
(2 \pi) \delta(k + k')
\frac{4 \pi}{q^2} \Delta(q,k) \\
&& \Delta(q,k) = \frac{q^4 G_v(q,k)^2}{4 \pi d^4 a^4 D_L^2(q,k)}
F(k)
\end{eqnarray}

Thus in the limit $q \to 0$ one obtains in general:

\begin{eqnarray}\label{Del}
\Delta(k) = \frac{1}{4 \pi} F(k) \frac{G_v(k)^2}{d^4 a^4 D_L^2(k)}
\end{eqnarray}
with $\Delta(k) = \Delta(0,k)$.
In the almost fully screened case of interest ($2\pi d\ll
a<\lambda$ or $2\pi d\gg a$) we have $G_v(k)/a^4 d^2 D_L(k)
\approx 1$, hence:

\begin{eqnarray}\label{Fk}
\Delta(k) = \frac{a^4}{4 \pi} F(k)\,.
\end{eqnarray}
For usual layered superconductors $2\pi d\ll a$ we have from Eqs.
(\ref{Geff1},\ref{T0}) for almost all $k$ ($k\gtrsim 1/a$)
$g(k)=2\beta T_0$, hence $\sigma_{eff}$ of Eq. (\ref{seffN}) with
$N=1$ becomes $\sigma_{eff}=4s$. Note that $\sigma_{eff}\sim B$,
hence defect formation is induced at a fixed disorder by
increasing the field $B$.

 We proceed now to study the full disorder problem on all
scales allowing for Bragg glass (BrG) properties
\cite{tgpldbragg}. The basic assumption is that the long range
extra displacement induced in the BrG configuration by the defect
is very small and one can expand in it. Consider then $H_{BG}(u)$
as the BrG hamiltonian for the $u$ field in presence of disorder
but in the absence of point defects. We add to it (here $u=u_L$):
\begin{eqnarray}
H(u) = H_{BG}(u) + \int_{q,k} h({\bf q},k) u({\bf q},k)\,.
\end{eqnarray}
In particular for the flux lattice problem we identify from Eq.
(\ref{Hvac1})
\begin{eqnarray}
h({\bf q},k) = \frac{1}{d^2 a^2} i q s({\bf q},k) G_v(q,k)\,.
\end{eqnarray}
The next order in the displacement expansion is $O(su^2)$ and
after integrating out $u_L({\bf q},k)$ leads to $s^3$ and higher order
terms; these are neglected in our low density treatment of
defects, i.e. large $\beta E_c$.

 Then one has the exact (although formal) expansion
for the free energy $F= - T \ln Z$:
\begin{eqnarray}\label{F}
F = F_{BG} +  \int_{q,k} h({\bf q},k) <u({\bf q},k)> - \frac{1}{2
T} \int_{q,q',k,k'} h({\bf q},k) h({\bf q}',k') <u({\bf q},k)
u({\bf q}',k')>_{c} + O(h^3)
\end{eqnarray}
where $<...>$ is thermal average in a particular disorder
configuration with no defects and $F_{BG}$ is the free energy of
the BG in that configuration and $c$ denotes connected averages; disorder
average will follow below.

In the absence of disorder the second term in Eq. (\ref{F}) is
zero and the third one yields the energy which screens the initial
defect-defect interaction:

\begin{eqnarray}
F_{screen} = - \frac{1}{2 d^2} \int_{q,k}
\frac{q^2 G_v(q,k)^2}{d^2 a^4 D(q,k)} |s({\bf q},k)|^2
\end{eqnarray}
using $<u({\bf q},k) u({\bf q}',k')>_c = T/D(q,k)$, i.e. the screening term in
Eq. (\ref{Geff}).

In presence of disorder, the disorder average of the third
term in Eq. (\ref{F}) still yields {\it exactly the same}
screening part of the interaction between defects. This is guaranteed
by the so-called statistical tilt symmetry of the Bragg glass
model in the absence of defects, i.e. the statistical invariance of
the disorder term in the Hamiltonian under
$u({\bf r},l) \to u({\bf r},l) + \phi({\bf r},l)$
where $\phi({\bf r},l)$ is an arbitrary function
(see e.g. \cite{tgpldbragg}) so that $\overline{
<u({\bf q},k) u({\bf q}',k')>_c}
\sim \delta^2F/\delta\phi^2|_{\phi=0}$ is independent of disorder.

Since this is an expansion in defect density $s({\bf q},k)$
we can now identify the random
potential coupling linearly to the defect via the second
term (a response of the third term in Eq. (\ref{F}) to defects
results in higher order $O(s^3)$ terms):
\begin{eqnarray}
V({\bf q},k) = - \frac{1}{d a^2}
G_v(q,k) i {\bf q} \cdot  <{\bf u}({\bf q},k)>_{BG}
\end{eqnarray}
The correlations are thus (overbar is disorder average):
\begin{eqnarray}
&& \overline{V({\bf q},k) V({\bf q}',k')}=
(2 \pi)^2 \delta({\bf q} + {\bf q}')
(2 \pi) \delta(k + k')  \frac{4 \pi}{q^2} \Delta(q,k) \\
&& \Delta(q,k) = \frac{1}{4 \pi d^2 a^4} q^4 G_v(q,k)^2
C_{BG}(q,k)
\end{eqnarray}
where $C_{BG}(q,k)$ denotes the disconnected average:

\begin{eqnarray}\label{uubar}
\overline{<{\bf u}({\bf q},k)> <{\bf u}({\bf q}',k')>} = (2 \pi)^2
\delta^2({\bf q} + {\bf q}') (2 \pi) \delta(k + k') C_{BG}(q,k)
\end{eqnarray}
At all temperatures except near melting one has:$ \overline{<u>
<u>} \approx \overline{<u u>}$ as thermal fluctuations are
subdominant.  Therefore we replace the left hand side of Eq.
(\ref{uubar}) by \cite{tgpldbragg}
\begin{equation}\label{CBG}
C_{BG}(q,k)\sim 1/({\bar q}^4+{\bar q}^3/R_c)
\end{equation}
where ${\bar q}^2= c_{66}q^2+c_{44}^Lk^2$ and $R_c$ is a Larkin
length along $c$. For $q=0$ and large $k\gtrsim 1/R_c$, i.e. on
short distances compared with $R_c$, this reduces to the previous
result Eq. (\ref{Fk}), while at longer scales the BrG induces
interlayer disorder correlation as seen by the defects. Replacing
$1/D_L^2(k)$ in Eq. (\ref{Del}) by $C_{BG}({\bf q},k)$ at $q=0$ we
obtain (using $G_v(k)/a^4 d^2 D_L(k) \approx 1$ as in Eq. \ref
{Fk})
\begin{equation}\label{Del2}
\Delta(k) = \frac{a^4}{4 \pi} F(k)\frac{k^4}{(k^4+R_c^{-1}k^3)}
\end{equation}

It is instructive to present another derivation of $\Delta (k)$,
valid at $T=0$. In general, the disorder potential $V({\bf r},l)$
couples to the flux density $\rho ({\bf r},{\bf u}({\bf r},l))$
and leads to a Bragg glass configuration ${\bf u}_{BG}({\bf
r},l)$. The addition of a vacancy at position ${\bf R}$ on layer
$l$ leads to an energy of $U({\bf R})=\sum_{l'}\int d^2r {\bf
u}_{vac}({\bf r}-{\bf R}, l'-l)\cdot {\mbox{\boldmath
$\nabla$}}\rho ({\bf r},{\bf u}_{BG}({\bf r},l'))V({\bf r},l')$.
One can now see that the force ${\mbox{\boldmath $\nabla$}}\rho
({\bf r},{\bf u}_{BG}({\bf r},l))V({\bf r},l)$ has short range
correlations. Indeed, at $T=0$ we can minimize the disorder energy
$\sum_{l}\int d^2r \rho ({\bf r},{\bf u}({\bf r},l))V({\bf r},l)$
with the elastic energy Eq. (\ref{Hel}) to yield ${\bf
u}_{BG}({\bf r},l))$, hence ${\mbox{\boldmath $\nabla$}}\rho ({\bf
r},{\bf u}_{BG}({\bf r},l))V({\bf r},l) \sim \nabla^2 {\bf
u}_{BG}({\bf r},l))$, the latter quantity having clearly short
range correlations $\sim {\bar q}^4/[{\bar q}^4+R_c^{-1}{\bar
q}^3]$. The potential $U({\bf R})$ is thus the convolution of a
short range correlated random force with the displacement ${\bf
u}_{vac}$ which has a long range form: for a single vacancy $|{\bf
u}_{vac}({\bf q},k)|^2\sim 1/q^2$ from Eq.\ (\ref{u}). Thus one
finds that $U(R)$ is logarithmically correlated with $\Delta (k)$
of Eq. (\ref{Del2}). Hence the BrG induces an effective disorder
correlation between layers on scales longer than $R_c$. For weak
disorder $R_c\gg d$ and the effect in $\int_k\Delta (k)$ is
negligible, hence the results of the Larkin regime are valid.

The application of our results to flux lattices depends on the
interlayer form of Eq.\ (\ref{Gv}) which for $a\gg d$ has the form
\cite{H3} $G_v(r,l)\sim e^{-ld/a}\ln r$, i.e. a range of
$l_0\approx a/d$. For usual layered superconductors \cite{Kes}
with $a/d\approx 10-100$, we find that the $N=1$ phase dominates
and $\sigma_{cr}=1/8$. The phase diagram has then the form of Fig.
2 with the magnetic field $B$ in the vertical axis

To achieve $N\neq 1$ phases the nearest layer coupling should
increase. We note that $g(k=0)=0$ since for a straight pancake rod
the logarithmic interaction is fully screened. Hence $\sum
_lJ_l=0$ and when the range $l_0$ is reduced $J_0, J_1$ dominate
the sum, i.e. $\eta_1 \rightarrow \frac{1}{2}$ when $d\gg a$, as
in Eq. (\ref{Geff11}). Direct evaluation of $\eta_1$ shows that it
crosses the critical value $1-1/\sqrt{2}$ when $d/a\approx 1$,
depending weakly on the ratio $a/\lambda_{ab}$. We therefore
propose that flux lattices in multilayer superconductors, where
$d>a$ can be achieved, may show a rich phase diagram with $N>1$
phases.

\section{Discussion}
We have developed here a variational method and a Cayley tree
rationale and applied these to the layered Coulomb gas. The
variational method is shown to reproduce the defect transition of
the single layer as well as demonstrate a first order transition
within the ordered phase. The latter was so far inferred in the
Caylee tree problem \cite{ct} or in the dynamic problem
\cite{dcpld}. To observe this transition one needs to induce
defects in the system, e.g. by finite size or dynamics. We also
show that this line survives in the disordered phase, showing a
crossover in the defect density dependence on temperature or
disorder.

The variational scheme has been extended to two layers, confirming
essentially the energy rationale. Near the onset to the $N=2$ rod
phase we find in a narrow interval a curious phase with a new
exponent relating the two components of the order parameter. We
consider then the variational scheme as reliable for the main
features of the phase diagram, i.e. the sequence of transitions
into rod phases (Fig. 1).

Our results are relevant to flux lattices where we find the phase
boundaries and propose that for $2\pi d\gtrsim a$ new $N>1$ phases
can be manifested. Our derivation assumes (i) dislocations are
neglected, and (ii) the Josephson coupling is neglected.
Assumption (i) implies that the melting transition is at higher
temperature or disorder than those of the defect transition. This
has been justified for the pure case in section VIA showing that
$T_{def}\ll T_m$ if either $2\pi d\ll a<\lambda$ or  $2\pi d\gg
a$. We assume that the same holds for disorder induced melting,
though the latter is less understood.

We discuss next assumption (ii), i.e. the effect of the interlayer
Josephson coupling J. In the absence of VI a layer decoupling was
found \cite{Daemen,H1} where $J$ vanishes on long scales. At this
transition
 the width of a Josephson flux line diverges and its fluctuations
 renormalize $J$ to zero. A complete description should allow for
 both VI defects and Josephson vortex loops which would combine to
 form 3-dimensional defect loops. We expect then that the defect
and decoupling transitions merge into one transition at $T_c$, above which
  both the renormalized $J$
 is zero as well as a finite VI density $n_{d}$ appears.

 In fact a transition to a "supersolid" phase in a flux lattice
in isotropic superconductors was proposed \cite{FNF} where a finite
 density of defect loops proliferate and a related
 "quartet" dislocation scenario was suggested \cite{feigelman}. In
 the supersolid desription a finite line energy competes with the
 entropy of the wandering line, both being linear in the defect
 length. The resulting transition temperature is comparable to
 that of melting \cite{FNF}, hence it is uncertain if this
 scenario is possible.

 In our VI transition the competing energies and entropies are
 logarithmic in the VI separation, rather than linear. If a Josephson
 coupling is added, naively a linear term is added since a flux
 line connecting the VI pair is formed. However, near decoupling
 the renormalized $J$ varies as  a power of scale, hence we expect that
 the free energy of a flux loop to be nonlinear in size, modifying
 significantly the supersolid transition at least in the small $J$
 case. We also show now that, in contrast with the supersolid
 scenario, $T_c$ can be well below melting.

 We note first that in
the pure system the decoupling transition is at $T_{dec}=8T_{def}$
(for $d\ll a\ll \lambda$) while its critical disorder (at $T=0$)
is at \cite{H1} $\sigma_{dec}=2=16\sigma_{def}$, hence the
$\sigma-T$ boundary of the defect transition is below that of
decoupling in both the $\sigma,T$ coordinates. The
disorder-temperature "phase diagram" has therefore 3 regions,
separated by the two lines $T_{def}(\sigma)$ and
$T_{dec}(\sigma)$: (i) decoupled and defected phase at high $T$ or
high $\sigma$, (ii) between the lines
$T_{def}(\sigma),\,T_{dec}(\sigma)$, and (iii) a coupled
defectless phase at small $T$ and small $\sigma$. This "phase
diagram" is inconsistent in the sense that $T_{def}(\sigma)$ is
derived in the absence of $J$, while $T_{dec}(\sigma)$ is derived
in the absence of VI defects.

We show next that $T_{def}<T_c<T_{dec}$. In phase (i)
$J\rightarrow 0$ and $n_d$ is relevant in the RG sense. This is
a consistent description since $J=0$ is assumed in the VI
description, hence region (i) is a disordered phase. In region
(iii) $n_d\rightarrow 0$ while $J$ is relevant, again a
consistent scenario since $J$ being relevant is shown assuming
$n_d=0$. However, in region (ii) both $n_d$ and $J$ are
relevant, hence seperate "decoupling" and "defect" descriptions
are inconsistent and a single combined transition within region
(ii) is expected, i.e $T_{def}<T_c<T_{dec}$. Since both
$T_{def},\,T_{dec}$ are well below melting for $a\ll \lambda$, we
conclude that $T_c$ is also well below melting.

In fact we can estimate $T_c$ by an argument as used in the $B=0$
case \cite{H3}. Consider the VI correlation length
$\xi_d\approx n_d^{-1/2}$ for $J=0$ (which diverges at
$T_{def}$) and the Josephson correlation length $\xi_J$ (which
diverges at $T_{dec}$). Consider a temperature for which
$\xi_J<\xi_d$; $\xi_J$ is the scale at which $J/T$ is
renormalized to strong coupling $\approx 1$, e.g. in 1st order RG \cite{H1}
\begin{equation}
\xi_J\approx a(T/J)^{1/[2(1-T/T_{dec}])}\,.
\end{equation}
 The Josephson term $J\cos
(\theta_{ns}+\theta_s)$ involves both the nonsingular phase
$\theta_{ns}$ and the singular one $\theta_s$ due to VI pairs. If
$\xi_J<\xi_d$ VI pairs are not seen on the scale between $a$
and $\xi_J$, renormalization of $J\cos (\theta_{ns})$ can proceed
till strong coupling is achieved, i.e. the phase is ordered. If
instead $\xi_J>\xi_d$ VI defects interfere in the $J$
renormalization and disorder the system. Hence $T_c$ is estimated
by $\xi_J\approx \xi_d$. From Eq. (\ref{sc3}) for the pure case,
\begin{equation}
\xi_d\approx a(e^{\beta E_c})^{1/[2(1-T_{def}/T)]}\,,
\end{equation}
hence $T_c$ is near $T_{def}$ if $J$ is sufficiently small,
\begin{equation}\label{Jcon}
J\ll Te^{-E_c/T}\,.
\end{equation}
For $Bi_2Sr_2CaCu_2O_8$  we estimate \cite{Kes,H3} $J\approx
 0.1 K$, $E_c\approx 10^3 K$ which for relevant $T=10-100 K$
does not satisfy Eq. (\ref {Jcon}), i.e. the transition is near
 $T_{dec}$. However, for multilayers such as
 $(Bi_2Sr_2CaCu_2O_8)_m(Bi_2Sr_2CuO_6)_n$, the semiconducting layers of
$Bi_2Sr_2CuO_6$ reduce $J$ by a factor $e^{-d/\xi}$ where
 $d\sim n$ includes now the thickness of the semiconducting layers. Hence, for
 a few such layers the condition (\ref{Jcon}) is already satisfied and
$T_c$ is near $T_{def}$.
Therefore, multilayer systems  are excellent candidates for observing the VI
transition with interlayer defects being either uncorrelated, when $2\pi d<a$,
or in
 correlated $N>1$  rod phases, when $2\pi d>a$. The latter condition is in fact
easier to realize in these multilayers where $d$ is larger.
 Increasing $d$ too much will push down the coupled phase to very
 low temperatures, hence the optimal case for study are multilayers
with $2\pi d\approx a$.

 Acknowledgements: This research was supported by THE ISRAEL
SCIENCE FOUNDATION founded by the Israel Academy of Sciences and
Humanities. B. H. thanks the Ecole Normale Sup{\'e}rieure for hospitality and
for support as visiting professor during part of this work.

\newpage
\appendix

\section{evaluation of some quantities}

Let us estimate:

\begin{eqnarray}
D_T(k) = \frac{1}{2} (\frac{1}{d a^2})^2
\sum_{{\bf Q} \neq {\bf 0}} (G_v(Q,k) - G_v(Q,0)) Q^2
\end{eqnarray}
in various regimes. Implicit in the sum is a cutoff at large $Q
\approx 2\pi/\xi$. One has:

\begin{eqnarray}
&& D_T(k) =  (\frac{1}{d a^2})^2  \frac{\phi_0^2 d^2}{8 \pi \lambda^2}
\sum_{{\bf Q} \neq {\bf 0}} (\frac{1}{1 + f(Q,k)} - \frac{1}{1 + f(Q,0)}) \\
&& = \frac{\phi_0^2}{8 \pi \lambda^2 a^4}
\sum_{{\bf Q} \neq {\bf 0}} (\frac{1}{1 + \frac{d}{4 \lambda^2 Q}
\frac{\sinh(Q d)}{\sinh^2(Q d/2) + \sin^2(kd/2)} }
- \frac{1}{1 + \frac{d}{4 \lambda^2 Q}
\frac{\sinh(Q d)}{\sinh^2(Q d/2)} ) }
\end{eqnarray}

Let us first consider the case $d < \xi_0$, $d/a \ll 1$. Then it simplifies
into:
\begin{eqnarray}
D_T(k) \approx \frac{\phi_0^2}{8 \pi \lambda^2 a^4} \sum_{{\bf Q}
\neq {\bf 0}} ( \frac{-\lambda^2 Q^2}{1 + \lambda^2 Q^2} +
\frac{\lambda^2 (Q^2  + k_z^2)}{1 + \lambda^2 (Q^2  + k_z^2)})
\end{eqnarray}
If we now further consider the case where $\lambda > a$ it
becomes:
\begin{eqnarray}
&& D_T(k) \approx \frac{\phi_0^2}{8 \pi \lambda^4 a^4} \sum_{{\bf
Q} \neq {\bf 0}} ( \frac{1}{Q^2} - \frac{1}{(Q^2  + k_z^2)})
\approx \frac{\phi_0^2}{8 \pi \lambda^4 a^4} (\frac{a}{2 \pi})^2
\pi \int_1^{a/\xi} dx (\frac{1}{x} - \frac{1}{x + (a k_z/2
\pi)^2})
\end{eqnarray}

Thus we find, for $d < \xi_0$, $d/a \ll 1$ and $\lambda > a$ that:
\begin{eqnarray}
D_T(k) \approx \frac{\phi_0^2}{32 \pi^2 \lambda^4 a^2}
\ln( \frac{1 + (k_z/Q_0)^2}{1 + (\xi_0 k_z/2 \pi)^2} )
\end{eqnarray}
where $Q_0=2\pi/a$ is the lowest term in the ${\bf Q}$ sum. In
general for $d>\xi$ one has:
\begin{eqnarray}
D_T(k) \approx \frac{\phi_0^2}{8 \pi \lambda^2 a^4} ( \sum_{{\bf
Q} \neq {\bf 0}}^{Q < 1/d} ( \frac{-\lambda^2 Q^2}{1 + \lambda^2
Q^2}+ \frac{\lambda^2 (Q^2  + k_z^2)}{1 + \lambda^2 (Q^2  +
k_z^2)}) +\sum_{{\bf Q} \neq {\bf 0}}^{Q > 1/d} ( \frac{1}{1 +
\frac{d}{2 \lambda^2 Q} \frac{1}{1 + 4 e^{- Q d} \sin^2(kd/2)}} -
\frac{1}{1 + \frac{d}{2 \lambda^2 Q}})
\end{eqnarray}

The first term can be estimated for $\lambda > a$ and the
second with no assumption:

\begin{eqnarray}
D_T(k) \approx \frac{\phi_0^2}{8 \pi \lambda^2 a^4}
[(\frac{1}{\lambda^2} \sum_{{\bf Q} \neq {\bf 0}}^{Q < 1/d}
(\frac{1}{Q^2} - \frac{1}{(Q^2  + k_z^2)}) + k_z^2 \sum_{{\bf Q}
\neq {\bf 0}}^{Q > 1/d} e^{- Q d} \frac{2 \lambda^2 Q d}{(1 +
\frac{2 \lambda^2}{d} Q)^2} )]
\end{eqnarray}

Using the previous calculation this yields:

\begin{eqnarray}
D_T(k) \approx \frac{\phi_0^2}{32 \pi^2 \lambda^4 a^2} \ln(
\frac{1 + (k_z/Q_0)^2}{1 + (d k_z)^2} ) \theta(d - a/(2 \pi)) +
\Phi(k)
\end{eqnarray}
and one can compute $\Phi(k)$ in two limits:

In the case $2 \pi d \ll a$ one finds:

\begin{eqnarray}
&& \Phi(k) \approx \frac{\phi_0^2}{2 Q_0^2 d^2 a^4} k_z^2 F[\frac{d}{\xi_0},
\frac{2 \lambda^2}{d^2} ] \\
&& F[x,y] = \int_1^x du \frac{u^2}{(1 + y u)^2} e^{-u}
\end{eqnarray}
Since $\lambda \gg d$ seems natural in that case one gets:

\begin{eqnarray}
&& \Phi(k) \approx \frac{\phi_0^2 d^2}{ 8 Q_0^2 a^4 \lambda^4}
k_z^2
\end{eqnarray}

In the opposite case $2 \pi d \gg a$ the sum is dominated by the
two shortest ${\bf Q}$ of length $Q_0$, hence

\begin{eqnarray}
D_T(k) \approx \Phi(k) \approx  \frac{\phi_0^2 Q_0 d}{2 \pi a^4}
k_z^2 e^{- Q_0 d} \frac{1}{(1 + \frac{2 \lambda^2}{d} Q_0)^2}\,.
\end{eqnarray}


\begin{thebibliography}{999}


\bibitem{tgpldbragg}
T. Giamarchi, P. Le Doussal, Phys. Rev. B {\bf 52} 1242 (1995) and
Phys. Rev. B {\bf 55} 6577 (1997).

\bibitem{Kes} P. H. Kes, J. Phys. I
France {\bf 6} 2327 (1996).

\bibitem{speakexp} B. Khaykovich et al. Phys. Rev. Lett {\bf 76}
2555 (1996), K. Deligiannis et al. Phys. Rev. Lett {\bf 79} 2121
(1997).

\bibitem{H2} B. Horovitz, cond-mat/9903167, Phys. Rev. B {\bf 60} R9939
(1999); cond-mat/0405016.

\bibitem{Dodgson} M. J. W. Dodgson, V. B. Geshkenbein and G.
Blatter Phys. Rev. Lett. {\bf 83} 5358 (1999).

\bibitem{GK} L.I. Glazman and A.E. Koshelev Physica {\bf 173 C} 180 (1991).

\bibitem{Daemen} L. L. Daemen, L. N. Bulaevskii, M. P. Maley and J. Y. Coulter,
Phys. Rev. Lett. {\bf 70}, 1167 (1993).

\bibitem{H1} B. Horovitz and T. R. Goldin, Phys. Rev. Lett. {\bf
80},1734 (1998); cond-mat/0405015.

\bibitem{H3} B. Horovitz, Phys. Rev. B {\bf 47}, 5947 (1993).

\bibitem{FNF} for the related putative transition to
a ''supersolid'' phase in non layered models see
E. Frey, D.R. Nelson and D.S. Fisher Phys. Rev. B {\bf 49} 9723 (1994).

\bibitem{HL} B. Horovitz and P. Le Doussal, Phys. Rev. Lett. {\bf 84}, 5395
(2000).

\bibitem{MHL} A. Morozov, B. Horovitz and P. Le Doussal,
cond-mat/0211080, Phys. Rev. B 67, 140505(R) (2003).

\bibitem{Bruynseraede} Y. Bruynseraede et al., Phys. Scr. {\bf T42},
37 (1992).

\bibitem{nattermann95} T. Nattermann et al., J. Phys. I (France) {\bf 5}, 565
(1995)

\bibitem{scheidl97} S. Scheidl, Phys. Rev. {\bf B 55}, 457 (1997)

\bibitem{tang96} L. H. Tang, Phys. Rev. {\bf B 54}, 3350 (1996).

\bibitem{dcpld} D. Carpentier and P. Le Doussal,
Phys. Rev. Lett. {\bf 81} 2558 (1998), cond-mat/9908335
Nuclear Physics B {\bf 588} 565 (2000), cond-mat/0003281
Physical Review E {\bf 63} 026110 (2001).

\bibitem{rdirac} B. Horovitz and P. Le Doussal, cond-mat/0108143,
Phys. Rev. B 65, 125323 (2002)

\bibitem{rem} B. Derrida Phys. Rev. B {\bf 24} 2613 (1981)

\bibitem{ct} B. Derrida, H. Spohn, J. Stat. Phys. {\bf 51} 817 (1988).

\bibitem{footnote0}
Note that a lattice model is used here as a mere convenience to
describe short scales, a continuum limit using hard-core charges
can also be constructed. The procedure, detailed in \cite{scheidl97,dcpld}
for the disordered case, yields a bare fugacity term
of the form $- \beta \sum_{{\bf r},l,a,l',b} E_{la,l'b} n_a({\bf r},l)
n_b({\bf r},l')$ with $E_{la,l'b} = \gamma K_{la,l'b}$,
corresponding a to a random bare core energy
$E^c_l({\bf r}) = E_c \pm v_l({\bf r})$ where
$v_l({\bf r})$ has short range correlations (over distance
$\sim \xi$).

\bibitem{footnote1} We note that \cite{hu,olive}
$E_c\approx (0.04-0.2)\tau$ while $J_0=2T_{def}$ in Eq. (\ref{T0})
yields $E_c/J_0 \approx (\lambda_{ab}/a)^2 /\ln(a/d)$; $\tau\,,
\lambda_{ab}\,,a$ are defined in section VI.

\bibitem{H4} B. Horovitz and A. Golub, Phys, Rev. B{\bf 55}, 14499
(1997); Phys. Rev. B{\bf 57}, 656(E) (1998).

\bibitem{scheidl98} S. Scheidl and M. Lehnen, Phys. Rev. B{\bf 58},
 8667 (1998).

\bibitem{hu} C. -R. Hu, Phys. Rev. B{\bf 6}, 1756 (1972); P. Minnhagen,
Rev. Mod. Phys. {\bf 59}, 1001 (1987).

\bibitem{olive} E. Olive and E. H. Brandt, Phys. Rev. B{\bf 57},
13861 (1998).

\bibitem{Goldin} T. R. Goldin and B. Horovitz, Phys. Rev. B{\bf
58}, 9524 (1998).

\bibitem{chamon96} C. C. Chamon, C. Mudry and X.G. Wen
Phys. Rev. Lett. {\bf 77} 4194 (1996).

\bibitem{korshunov_nattermann_diagphas}
S. E. Korshunov and T. Nattermann, Phys. Rev. B {\bf 53} 2746
(1996).

\bibitem{castillo} H. E. Castillo, C. C. Chamon, E. Fradkin,
P. M. Goldbart and C. Mudry, Phys. Rev. B {\bf 56}, 10668 (1997)

\bibitem{freezing}
H. E. Castillo and P. Le Doussal, Phys. Rev. Lett. {\bf 86} 4859
(2001).

\bibitem{feigelman}
M. V. Feifel'man, V. B. Geshkenbein and A. I. Larkin, Physica C{\bf 167},
177 (1990).

\end{thebibliography}

\end{document}